\documentclass{aa}  
\usepackage{amsmath}
\usepackage{graphicx}
\usepackage{txfonts}
\usepackage{natbib}
\usepackage{psfrag}
\newcommand{\ltapprox}{\raisebox{-0.5ex}{$\,\stackrel{<}{\scriptstyle\sim}\,$}}
\newcommand{\gtapprox}{\raisebox{-0.5ex}{$\,\stackrel{>}{\scriptstyle\sim}\,$}}
\newcommand{\lya}{\ifmmode {\rm Ly}\alpha \else Ly$\alpha$\fi}

\def\msun{\ifmmode M_{\odot} \else M$_{\odot}$\fi}
\def\zsun{\ifmmode Z_{\odot} \else Z$_{\odot}$\fi}
\def\lsun{\ifmmode L_{\odot} \else L$_{\odot}$\fi}

\begin{document}
   \title{Looking for the first galaxies: Lensing or blank fields? }

   \author{A. Maizy
          \inst{1}
          \and
J. Richard\inst{2}
          \and
M. A. De Leo\inst{3}
          \and
R. Pell\'o\inst{1}
          \and
J. P. Kneib\inst{4}
          }

   \offprints{A. Maizy}
   \institute{
Laboratoire d'Astrophysique de Toulouse-Tarbes, 
CNRS, Universit\'e de Toulouse, 14 Av. Edouard-Belin, F-31400 Toulouse, France\\
              \email{alexandre.maizy@ast.obs-mip.fr}
              \email{roser@ast.obs-mip.fr}
         \and
Institute for Computational Cosmology,
Department of Physics, Durham University,
    South Road, Durham, DH1 3LE, UK  \\
              \email{johan.richard@durham.ac.uk}
         \and
Instituto de Astronom\' \i a, UNAM, Apartado Postal 70-264, 04510 M\'exico DF,
Mexico \\
              \email{madeleo@astroscu.unam.mx}
         \and
OAMP, Laboratoire d'Astrophysique de Marseille, UMR 6110 traverse du Siphon, 13012 Marseille, France \\
              \email{jean-paul.kneib@oamp.fr}
}

   \date{Received ; accepted }


\abstract
   {The identification and study of the first galaxies remains one of the most
exciting topics in observational cosmology. The determination of the best
possible observing strategies is a very important choice in order to
build up a representative sample of spectroscopically confirmed sources
at high-z ($z\gtapprox7$), beyond the limits of present-day
observations. 
}
   {This paper is intended to precisely adress the relative efficiency of
lensing and blank fields in the identification and study of galaxies at
$6\ltapprox z \ltapprox 12$. 
}
   {The detection efficiency and field-to-field variance are estimated from
direct simulations of both blank and lensing fields observations. Present
known luminosity functions in the UV are used to determine the expected
distribution and properties of distant samples at $z\gtapprox6$ for a
variety of survey configurations. Different models for well known lensing
clusters are used to simulate in details the magnification and dilution
effects on the backgound distant population of galaxies.  
}
   {The presence of a strong-lensing cluster along the line of sight has a 
dramatic effect on the number of observed sources, with a positive
magnification bias in typical ground-based ``shallow'' surveys
($AB\ltapprox25.5$). The positive magnification bias increases with the 
redshift of sources and decreases with both depth of the survey and the
size of the surveyed area. The maximum efficiency is reached for lensing
clusters at $z\sim0.1-0.3$. Observing blank fields in shallow surveys is
particularly inefficient as compared to lensing fields if the UV LF for
LBGs is strongly evolving at $z\gtapprox7$. Also in this case, the number
of $z\ge8$ sources expected at the typical depth of JWST ($AB\sim28-29$)
is much higher in lensing than in blank fields (e.g. a factor of $\sim10$
for $AB\ltapprox28$).
{All these results have been obtained assuming that number counts derived
in clusters are not dominated by sources below the limiting surface brightness
of observations, which in turn depends on the reliability of the usual
scalings applied to the size of high-z sources. }
}
   {Blank field surveys with a large field of view are needed to prove the
bright end of the LF at $z\gtapprox6-7$, whereas lensing clusters are
particularly useful for exploring the mid to faint end of the LF.    
}


   \maketitle
%


\section{\label{intro} Introduction}

   Constraining the abundance of $z>7$ sources remains an important challenge
of modern cosmology. Recent {\tt WMAP} results place the first building
blocks of galaxies at redshifts $z=11.0 \pm 1.4$, suggesting that reionization was an extended process 
\citep{Dunkley09}. Distant
star-forming systems could have been responsible for a significant part of the
cosmic reionization. Considerable advances have been made during the last
years in the observation of the early Universe with
the discovery of galaxies at $\sim$6-7, close to the end of the
reionization epoch (e.g. \citealt{Hu,Kodaira,Cuby03,Kneib04,Stanway04,Bouwens04a,Bouwens06,Iye06,Bradley,Zheng}),
and the first prospects up to z$\sim$10 (\citealt{Pello04,Richard06,Stark07,Richard08,Bouwens08,Bouwens09}).

  High-$z$ surveys are mainly based on two different and complementary
techniques: the dropout (Lyman-$\alpha$ Break) identification, which is an
extrapolation of the drop-out technique used for Lyman-Break Galaxies (LBGs, \citealt{Steidel99})
to higher redshifts (e.g. \citealt{Bouwens06,Richard06,Richard08}),
and the narrow-band (NB) imaging aimed at detecting Lyman $\alpha$
emitters (LAEs, e.g. \citealt{Taniguchi,Iye06,Kashikawa,Willis_ZEN,Cuby07}). 
Using the former technique, \citet{Bouwens08} found a
strong evolution in the abundance of galaxies between z$\sim$7-8 and
z$\sim$3-4, the SFR density beeing much smaller at very high-z up to the
limits of their survey, in particular towards the bright end of the Luminosity
Function (LF). A strong evolution is also observed in the number density of
sources detected with NB techniques, which seems to be much smaller at $z\ge7$
than in the z$\sim 5-7$ interval (\citealt{Iye06,Cuby07,Willis_ZEN3}).  

   Both dropout and NB approaches require a subsequent
spectroscopic confirmation of the selected candidates. 
For now approximately ten galaxies beyond z$\sim$6.5 are known
with secure spectroscopic redshifts (\citealt{Hu,Kodaira,Cuby03,Taniguchi,
Iye06}). All samples beyond this redshift are mainly supported by photometric 
considerations (\citealt{Kneib04,Bouwens04b,Bouwens06,Richard06,Richard08,Bradley}).
This situation is expected to dramatically improve in the near future with the
arrival of a new generation of multi-object spectrographs in the near-IR, such as
MOIRCS/Subaru, Flamingos2/Gemini-S ($\sim$ 2009), or EMIR/GTC\footnote{ 
http://www.ucm.es/info/emir/} ($\sim$ 2012), 
with well suited field of view,
spectral resolution and sensitivity. These forthcoming facilities should provide 
spectroscopic confirmation for a large number of z$>$7 candidates identified
from deep photometric surveys, and the first characterization of the physical
properties of these sources (e.g. IMF, stellar populations, fraction of AGN,
...). 

   The aim of this paper is to determine the best possible observing strategies
in order to build up a representative sample of spectroscopically confirmed
$z\ge7$ galaxies. The photometric pre-selection of high-$z$ candidates could
be achieved either in blank fields or in lensing clusters. This later technique, also
first referred to as the ``gravitational telescope'' by Zwicky , has proven highly
successful in identifying a large fraction of the most distant galaxies known
today thanks to magnifications by typically 1-3 magnitudes in the cluster core 
(e.g. \citealt{Ellis01,Hu,Kneib04,Bradley,Zheng,Bradac09}). 
The presence of a strong
lensing cluster in the surveyed field introduces two opposite effects on
number counts as compared to blank fields. In one hand, gravitational magnification
increases the number of faint sources by improving the detection towards the
faint end of the LF. On the other hand, the reduction of
the effective surface by the same magnification factor leads to a dilution in
observed counts. The global positive/negative magnification bias obviously
depends on the slope of the number counts, as pointed out by \citet{Broadhurst}. 

   This paper addresses the relative efficiency of surveys conducted on blank
and lensing fields as a function of the relevant parameters, namely the
redshift of the sources, the redshift and properties of the lensing
clusters and the survey characteristics (i.e. area, photometric depth...). 
This calculation requires a detailed simulation of observations using lensing
models, and realistic assumptions for the properties of background
sources according to present-day observational results, in particular for the luminosity 
function and typical sizes of $z>7$ galaxies.  

   The paper is organized as follows. In Section \ref{method} we describe the
simulations performed in order to determine the relative detection efficiency
for high-$z$ sources, both in lensing and blank fields. Section \ref{results}
presents the results, in particular the detection efficiency achieved as a
function of redshift for both sources and lensing clusters, together with a
discussion on the influence of lensing cluster properties and field-to-field
variance. A discussion is presented in Section \ref{discussion} on the
relative efficiency as a function of survey parameters, and a comparison
between simulations and present surveys. Conclusions are given in Section
\ref{conclusions}.  

   Throughout this paper, we adopt a concordance cosmological model, with 
$\Omega_{\Lambda}=0.7$, $\Omega_{m}=0.3$, and $H_{0}=70\ {\rm km}\ {\rm s}^{-1}\
Mpc^{-1}$. All magnitudes are given in the AB system. Conversion values
between Vega and AB systems for the filters used 
in this paper are typically C$_{AB}$= 0.95, 1.41 and 1.87 in $J$, $H$ and $K$
respectively, with $m_{AB}=m_{Vega}+C_{AB}$.


\section{\label{method} Simulations of lensing and blank field observations}

\subsection{Simulation parameters}
This section describes the ingredients used in the simulations to implement different
assumptions that would affect our efficiency in detecting high redshift
galaxies. There are three important aspects to be considered in the comparison
between lensing and blank fields. The first one is the LF and typical sizes of
sources. The second one concerns the properties of the lensing clusters, in
particular their mass distribution and redshift. The third one is related to
the survey parameters, namely the photometric depth and the size of the
field. All these aspects  
are discussed in this section. Table~\ref{surveyparam} provides the list of
parameters used in these simulations, together with the range of values
explored. 

\subsubsection{\label{sources} Source Properties}

These simulations are focused on the detection of sources in the redshift
range $6<z<12$, a relevant domain for  
spectroscopic follow-up with near-infrared instruments. The lower limit of
this redshift domain overlaps with current photometric  
surveys measuring the LF at $z\sim6-7$ (e.g. \citealt{Bouwens07}). However,
the LF is still largely unconstrained beyond $z\gtapprox7$ because of the lack
of spectroscopic confirmation of photometric samples and the relatively small
size of the surveyed volumes.  

The abundance of background sources at these redshifts is given by the luminosity 
function $\phi(L)$, with L the rest-frame UV luminosity at 1500 \AA. $\phi(L)$ is the most basic description 
of the galaxy population from an observer point of view. We adopt a
parametrization based on the analytical Schechter  
function \citep{Schechter76}:  

\begin{equation}
\phi(L)\ dL  =  \phi^* \left(\frac{L}{L^*}\right)^{-\alpha}
\exp \left(-\frac{L}{L^*}\right)\, d\!\left(\frac{L}{L^*}\right)
\end{equation}

The slope at faint luminosities $\alpha$, the characteristic luminosity $L^*$
and the normalization factor $\Phi^*$ have been constrained by several
photometric surveys targeting Lyman Break Galaxies (LBG) at high redshift
($z\ge4$) (\citealt{Ouchi04}, \citealt{Beckwith06}, \citealt{Bouwens07},
\citealt{Bouwens08}, \citealt{McLure09}, \citealt{Henry09}).    
Three different representative cases are being discussed in our simulations,
basically exploring our present knowledge (or lack of knowledge) of
the overall shape of the LF at $z\gtrsim6$ (Table~\ref{surveyparam}): 

\begin{itemize}
\item[(a)]{An "optimistic" scenario where LBGs show no-evolution from $z\sim6$,
with the LF parameters as determined by \citet{Beckwith06}. Indeed, the LF at
$z\sim6$ found by these authors display the same shape as for $z\sim3$
(\citealt{Steidel99}), but a 3 times smaller normalization factor (but see,
e.g., \citealt{McLure09}).}

\item[(b)]{A constant LF based on robust measurements by \citet{Bouwens07} at $z\sim6$ 
in the Hubble UDF, but using the more recent fit parameters from
\citet{Bouwens08}. As compared to model (a), this LF exhibits a turnover
towards the bright end.}

\item[(c)]{The evolutionary LF recently proposed by \citet{Bouwens08}, which includes an 
important dimming of $L^{*}$ with increasing redshift. This LF represents the ``pessimistic'' case 
with respect to the case $(a)$, with very few high-luminosity
galaxies.}
\end{itemize}

The size of the sources is a relevant parameter in this study, given the
finite resolution of instruments, and the fact that gravitational
magnification preserves surface brightness. High redshift galaxies are
expected to be very small at $z>7$, typically $\ltapprox 0.``10$ on the sky
($e.g.$ \citealt{Barkana00}). 
Recent observations of photometric candidates
support this idea (\citealt{Bouwens08},
\citealt{Oesch09}). 
This means that a large fraction of
lensed sources should remain spatially unresolved in ground-based surveys,
even with a strong gravitational magnification (hereafter $\mu$) of
$\sim10$. The high resolution capability of JWST is clearly needed for
resolving such faint galaxies. In the present simulations and for detection
purposes, we consider all sources at $z>6$ as spatially unresolved. 
However, galaxy morphology and image sampling are important when
discussing the efficiency of surveys based on space facilities, 
as discussed in Sect.~\ref{size}.

\subsubsection{\label{lenstool} Lensing effects}

\begin{figure*}
\centering

\begin{tabular}{c c c}
\includegraphics[width=6cm]{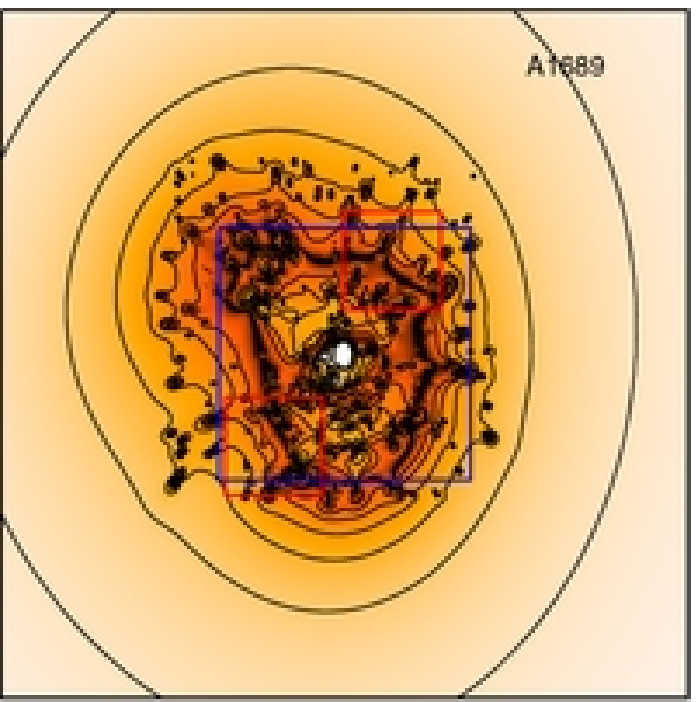} 
\includegraphics[width=6cm]{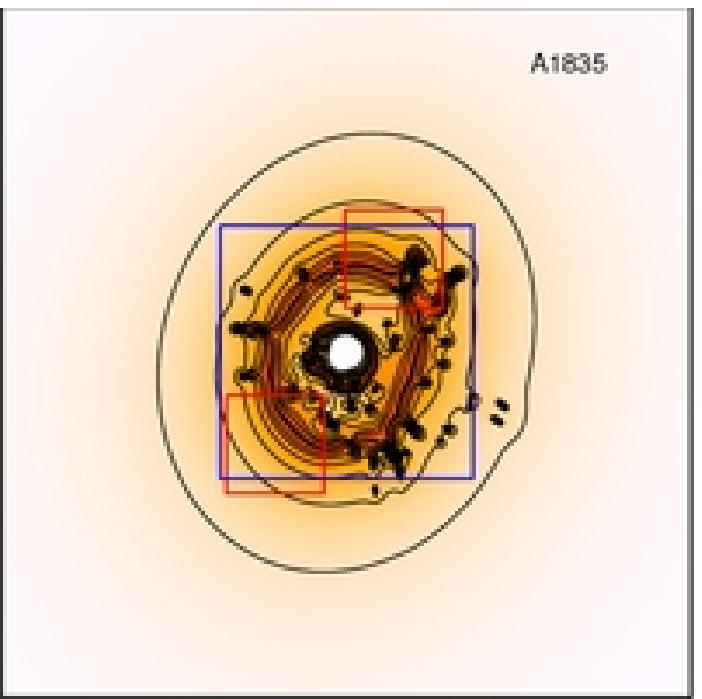} 
\includegraphics[width=6cm]{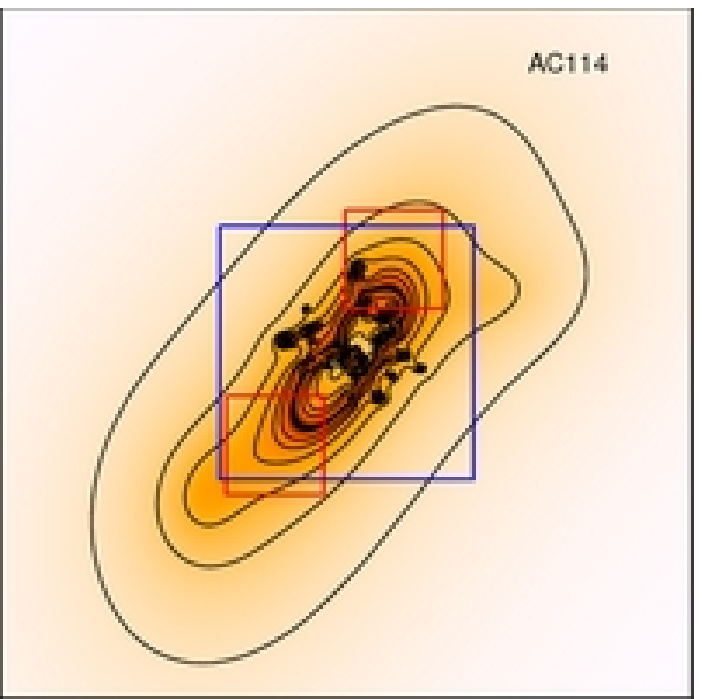} \\
\end{tabular}

\caption{Magnification maps for the three clusters at $z_s=8$ used in this
  study (from left to right A1689, A1835 and AC114 respectively). The global
  size of the image corresponds to the $6'\times6'$ FOV whereas the blue and
  red squares correspond respectively 
  to $2.2'\times2.2'$ and $0.85'\times0.85'$ FOV
  (see Sect.~\ref{simulation}). Black contours represent different
  magnification regimes with increasing magnification values from 0.5 to 3
  mags towards the cluster center. \label{cluster}} 

\end{figure*}

\begin{figure}
 \centering
 \includegraphics[width=9cm]{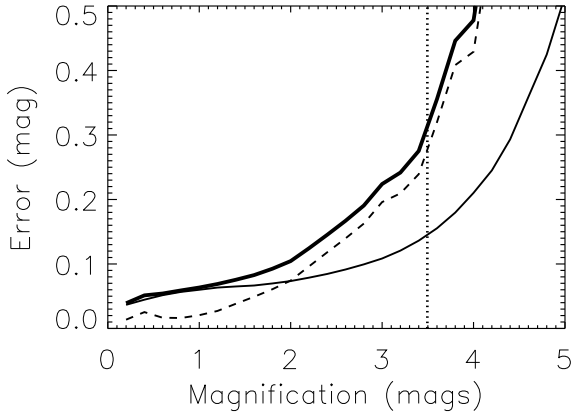}
\caption{Typical error in the magnification factor $\mu$, as a function of the magnification for Abell 1835. 
The solid curve gives the statistical error derived from the MCMC model, while the dashed
curve gives the systematic error between two choices of parametrization (see text for details).
The thick solid curve is the quadratic sum of both errors, used later in the calculation. 
The vertical line represents the conservative upper limit of $\mu$=25. \label{magerr}}

\end{figure}

The present simulations address the effect of lensing by a foreground galaxy
cluster. Several well-studied examples of lensing clusters are used in order
to evaluate the influence of different mass distributions on the final
results. Reference lensing clusters usually display several multiply-imaged
systems with redshift measurements, allowing us to model their lensing
properties accurately. Lensing clusters considered in these simulations have
been previously used as gravitational telescopes to search for high redshift
dropouts and LAEs. 
We take advantage of this situation to perform a direct comparison between our
estimates and available observations. Finally,
we selected clusters with different redshifts, total mass distributions and 
morphologies, because all of these factors are susceptible to affect the way
they magnify background galaxies. 

We selected three clusters satisfying the previous criteria: Abell 1689, Abell 1835 and AC114. 
Abell 1689 is one of the most spectacular gravitational telescopes, having the 
largest Einstein radius observed to date (45''). Both optical dropouts
\citep{Bradley} and Lyman-$\alpha$ emitters  
\citep{Stark07} candidates have been reported in the background of this cluster.
Abell 1835 and AC114 are both massive, X-ray luminous clusters, previously
used in our deep near-infrared survey for high redshift dropouts with
VLT/ISAAC \citep{Richard06}. Finally, these three clusters constitute the  
sample used by the ZEN2 survey for LAEs in narrow-band images \citep{Willis_ZEN3}.

We used the most recent mass models available for the reference clusters to
derive the magnification maps (see Table \ref{surveyparam}) although
simulation results are found to be weakly sensitive to modeling details. 
Each lensing cluster has been modeled in a similar way using the public
lensing software Lenstool\footnote{{\tt
http://www.oamp.fr/cosmology/lenstool}}, including the new MCMC
optimization method \citep{Jullo} providing bayesian estimates on each
parameter derived from the model.  

The structure of mass models is given by a sum of individual dark matter
subcomponents of two different types: large scale components, reproducing the
cluster-scale behavior of dark matter, and small scale potentials centered on
each cluster galaxy, reproducing the effect of substructure. Each lensing
potential is parametrized by a Pseudo-Isothermal Elliptical Mass Distribution
model (PIEMD, \citealt{Kassiola}), with a projected mass density $\Sigma$
given by:  

\begin{equation}
\Sigma(x,y)=\frac{\sigma_0^2}{2G}\ \frac{r_{cut}}{r_{cut}-r_{core}}\ \Big[\frac{1}{(r_{core}^{2}+\rho^2)^{1/2}}-\frac{1}{(r_{cut}^2+\rho^2)^{1/2}} 
  \Big], 
\end{equation}

where $\rho^2=[(x-x_c)/(1+\epsilon)]^2+[(y-y_c)/(1-\epsilon)]^2$, ($x_c$,
$y_c$) stands for the central position with respect to the BCG (cD or central
Bright Cluster Galaxy), $\epsilon=(a-b)/(a+b)$ is the ellipticity, $\sigma_0$
is the central velocity dispersion and ($r_{core}$,$r_{cut}$) are two
characteristic radii. For each lensing potential, the position of $x$ and $y$
axis is given by the position angle $\theta$. The total mass of this profile
is proportional to $r_{cut}\ \sigma_0^{2}$. 
The PIEMD parametrization, easily linked to the observed geometry of
elliptical lensing galaxies, has been widely used to  
model the strong lensing properties of massive clusters \citep{Smith05, 
Richard07, Limousin}.  

A good approximation of the angular distance of the critical line,
corresponding to maximum magnification in a flat universe, is given by the
Einstein radius $\theta_E$: 

\begin{equation}
\theta_E=\frac{4\pi\ \sigma_0^2}{c^2} \frac{D_A(z_s)-D_A(z_{c})}{D_A(z_s)}
\label{Einstein}
\end{equation}

where $D_A(z)$ stands for the angular distance at redshift $z$. Source and
cluster redshifts are, respectively, $z_s$ and $z_{c}$.  

The value of $\theta_E$ provides a fair estimate of the extension of the
strongly magnified area ($\mu>10$) in the image plane. This value quantifies
the power of a gravitational telescope to magnify background sources. Equation
\ref{Einstein} shows that, for a given source redshift $z_s$, $\theta_E$
depends on $\sigma_0^{2}$ and on the cluster redshift $z_{c}$.  

For the three clusters mentioned before, there is a significant variation in redshift ($z_{c0}\sim0.17-0.3$) and 
in $\sigma_0$ (taken from the mass models and reported in Table \ref{surveyparam}), Abell 1689 being $\sim30\%$ more
massive and less distant than AC114, for instance. We explored a wider range of cluster redshifts in our simulations, 
producing fiducial lensing clusters by adjusting $z_{c}$ between 0.1 and 0.8 in the three cases, assuming 
no evolution in the cluster properties. This is clearly an over-simplistic and
conservative assumption, as massive clusters of galaxies undergo a dynamical
evolution during cluster assembly at high redshift. 

The relevance of the MCMC approach of the cluster mass modeling is to derive relevant 
statistical errors in the magnification factors. Fig.~\ref{magerr} illustrates the 
typical errors in the magnification at a given position of the lensing field, 
in the case of the lensing cluster Abell 1835. Similar errors in the magnification 
are found in the case of the two other clusters, as all of them have $\ge5$ multiple systems constraining
independent regions of the cluster cores, the majority of them having
spectroscopic redshifts. For reasonable magnification factors 
($\le 3$ magnitudes), this error is always smaller than 0.1 magnitudes (or $\sim10\%$ relative
error in flux). For larger magnifications factors, corresponding
to the vicinity ($\le 2''$) of the critical lines, the error
can reach much higher values. The \textit{systematic errors} 
in the magnification factors, due to the choice of the parametrization when 
building the lensing model, can be estimated for Abell 1835, which have been 
modelled by Richard et al.(2009, submitted) using both PIEMD profiles and 
Navarro-Frenk-White (NFW, \citealt{Navarro}) profiles for the cluster-scale 
mass distributions. The comparison of magnifications from both models, at a given
position, gives an estimate of the systematic error in the magnification, which 
dominate at large $\mu$ (Figure~\ref{magerr}), reaching typical values of 0.3 magnitudes. 
We adopted a conservative upper limit of $\mu=25$ to avoid singularities in the magnification
determination. This is justified by the finite resolution of
instruments, and the limited knowledge on the precise location
of the critical lines at such high z (typically $\sim1''$). The affected area is not 
significant once averaged over the entire field of view. 
Nevertheless, the quadratic sum of the statistical and systematic errors in the magnification 
is later used to derive errors in the number density calculations when looking 
at lensing fields.

\begin{table*}[ht]
\begin{tabular}{llllll}
Parameter & Explored range &&&& Reference\\
\hline
LF (a) & ($<z>=6.0$)   & $\alpha=1.6$,   & $\phi^*=0.4\ 10^{-3}$ Mpc$^{-3}$, & $M^*=-21.07$ & \citet{Beckwith06}  \\
LF (b) & ($<z>=5.9$)   & $\alpha=1.74$, & $\phi^*=1.1\ 10^{-3}$ Mpc$^{-3}$, & $M^*=-20.24$ & \citet{Bouwens08}\\
LF (c) & ($3.8<z<7.4$)& $\alpha=1.74$, & $\phi^*=1.1\ 10^{-3}$ Mpc$^{-3}$, & $M^*=-21.02+0.36$\ $(z-3.8)$ & \citet{Bouwens08}\\
Source redshift& $ z_{s}=6-12$ &&&& \\
\hline
Lensing cluster & Abell 1689  &$z_{c0}=0.178$, &$\sigma_0$=1320 km/s, & $N_{\rm sub}$=266& \citet{Limousin}\\
                             & Abell 1835  &$z_{cl0}=0.25$,   &$\sigma_0$=1210 km/s, & $N_{\rm sub}$=90& Smith(2005) Richard(2009) \\
                             & AC114         &$z_{cl0}=0.31$,   &$\sigma_0$=1080 km/s, & $N_{\rm sub}$=28& \citet{Natarajan98} \\
Cluster redshift $z_{c}$ & $0.1-0.8$  &&&& \\
\hline
Survey strategy & GTC/EMIR &FOV=$6'\times6'$, &pix: $0.2''/pixel$, &depth: $\Delta z=1$ & \citet{Garzon}\\
& JWST/NIRCam &FOV=$2.2'\times2.2'$, &pix: $0.06''/pixel$, &depth: $\Delta z=1$ & \citet{Rieke} \\
& HST/NICMOS &FOV=$0.85'\times0.85'$, &pix: $0.2''/pixel$, &depth: $\Delta z=1$ &  \citet{Thompson} \\
\end{tabular}
\caption{\label{surveyparam} Summary of the parameters included in our
simulations. For each entry, we give the range of values explored and
reference to the relevant publication. 
LF(a) is the same as in \citet{Steidel99}, but $\Phi^*$  is a factor of 3
  smaller (see Sect.~\ref{sources}).
}
\end{table*}

\subsubsection{\label{simulation} Survey Simulations}

In addition to cluster and source properties, the main ingredients to consider in the simulations are the following: 
\begin{itemize}
\item The typical field of view (FOV) of near-IR instruments for 8-10 meters
  class telescopes and space facilities. The former typically range between a
  few and $10'$ on a side (e.g. $\sim6'\times6'$ for EMIR/GTC,
  $\sim7'\times7'$ for Hawk-I/VLT). The later are usually smaller
  (e.g. $\sim1'\times1'$ for NICMOS/HST, $\sim2.2'\times2.2'$ for JWST or
  WFC3-IR/HST). Fig.~\ref{cluster} presents the comparison between these
  typical FOV values and the magnification regimes found in lensing
  clusters. The references for the different instruments used in the
  simulations are presented in Table ~\ref{surveyparam}. 
\item The limiting magnitudes of present near-IR surveys based on ground-based
  and space observations tailored to select LBGs at $z\ge6$. The former are
  typically limited to $AB\sim25.5$ (see Sect. \ref{sources}), whereas the
  later could reach as deep as $AB\sim29$ with JWST (see Sect.~\ref{variance}
  and ~\ref{discussion}).  
\end{itemize}


The shallow magnitude limit of $AB\sim25.5$ achieved on ground-based
observations should allow us to detect galaxies with a UV continuum
corresponding to a SFR $\sim40/\mu$ $\msun/yr$ at $z\sim10$, whereas the
typical depth for JWST should be $\sim1/\mu$ $\msun/yr$. 

We can relate the UV luminosities of high redshift galaxies with the expected Lyman-$\alpha$ 
emission line by converting $L$ into a star formation rate SFR using the calibrations from 
\citet{Kennicutt}:

\begin{equation}
{\rm SFR} (M_\odot\ {\rm yr}^{-1})= 1.4\ 10^{-28}\ L\ ({\rm ergs\ Hz^{-1}\ {s}^{-1}})\\
\end{equation} 

The expected Lyman-$\alpha$ luminosity produced at the equilibrium can be written as:

\begin{equation}
L_{Ly_\alpha} ({\rm ergs\ s^{-1}}) = (1-f_{esc})\ f_\alpha\ {\rm SFR} (M_\odot\ {\rm yr}^{-1})\\
\end{equation} 

where $f_{esc}$ is the escape fraction of Lyman$-\alpha$ photons and $f_\alpha$ the Lyman-$\alpha$ production rate 
per unit of star formation. Assuming no reddening, the typical values for
$f_\alpha$ range between 2.44 10$^{42}$ and 6.80\ $10^{42}$ $ergs.s^{-1}$
\citep{Schaerer02}. We use these scaling relations when discussing the
detectability of Lyman-$\alpha$ in lensing fields. 

\subsection{\label{implementation} Implementation}

We explicitly compute the expected number counts $N(z,m_0)$ of sources at the
redshift z brighter than a limiting magnitude $m_0$ by a pixel-to-pixel
integration of the (magnified) source plane as a function of redshift, using
the sources and lensing cluster models described in the previous
subsections. 
Number counts are integrated hereafter within a redshift slice $\Delta
  z=1$ around z, unless otherwise indicated. 
With respect to a blank
field, the magnification pushes the limit of integration to fainter
magnitudes, whereas the dilution effect reduces the effective volume by the
same factor.  

An important effect to take into account in cluster fields is light
contamination coming from the large number of bright cluster galaxies, which
reduces the surface area reaching the maximum depth, and consequently prevents
the detection of faint objects, especially in the vicinity of the cluster
center. This contamination effect can be as high as 20\% of the total surface
\citep{Richard06}, whereas it is almost negligible in blank field surveys.  

\begin{table*}[ht]
\centering
\begin{tabular}{llllllll}
Cluster & Instrument and Reference & Filter & Exposure Time & Depth
(5$\sigma$, AB) & Mask area $6'\times6'$, & $2.2'\times2.2'$, &
$0.85'\times0.85'$ \\ 
\hline
Abell 1689 & ISAAC/VLT \citet{Moorwood97} & SZ & 14.4 ksec & 26.0 & 7\% & 12\% & 10\%   \\
Abell 1835 & FORS/VLT  \citet{Appenzeller} & V  & 19.6 ksec & 27.0 & 6\% & 10\% & 8\%   \\
AC114      & SOFI/NTT  \citet{Moorwood98} & K  & 10.8 ksec & 22.0 & 6\% & 9\% & 7\%   \\
\end{tabular}
\caption{\label{masks} Characteristics of the images used to produce the
  foreground object's mask for each cluster field, the three last columns
  representing the fractional surface area covered by the foreground galaxies
  in each cluster, for the three reference field of view, EMIR-like, JWST-like
  and NICMOS-like respectively} 
\end{table*}

We created bright-objects masks by measuring the central position
($x_c$,$y_c$) and shape parameters ($a$, $b$, $\theta$) of galaxies in the
three cluster fields, each object being approximated by an ellipse during this
process. We used SExtractor \citep{SExtractor} in combination with reasonably
deep and wide ground-based images available from the ESO archive (larger than
6' $\times$ 6' used in these simulations). The characteristics of these images
are summarized in Table \ref{masks}. They were reduced using standard IRAF
routines. The image mask $M(x,y)$ produced is the superposition of ellipses
for all objects in the photometric catalog where pixels belonging to object
domains were flagged. Ellipses correspond to $1-\sigma$ isophotes over the
background sky. In other words, only images lying on empty regions have been
included in the lensed samples, thus providing a lower limit for detections in
lensing fields.This fractional area covered by foreground galaxies ranges
between 6\% and 12\% depending on the cluster as well as the size and central
position of the field of view (Table \ref{masks}). The largest hidden area
corresponds to the smaller field of view centered on the cluster
(JWST-like). NICMOS pointings are even smaller, but they are centered on the
critical lines in our study and avoid the crowded central regions of the
cluster. In blank fields, this value doesn't exceed 3-4\%.  
In the next sections, we have taken into account this correction in the
calculations of number counts, both in blank fields and lensing fields. 

Including the object mask $M(x,y)$, number counts $N(z,m_0)$ are given by the following expression:

\begin{equation}\label{e:barwq}\begin{split}
N(z,m_0)&=\phi^{*} \int\limits_{x,y} M(x,y) \int\limits_{L(\mu,z,m_0)}^{\infty} \frac{Cv(x,y,z)}{\mu(x,y,z)} \left(\frac{L(\mu,z,m_0)}{L^*}\right)^\alpha\\
&\quad\cdot exp\left(-\frac{L(\mu,z,m_0)}{L^*}\right)\, d\!\left(\frac{L}{L^*}\right)\ dx\ dy
\end{split}\end{equation}

where $\mu(x,y,z)$ is the magnification induced by the lensing field,
$C_{v}(x,y,z)$ is the covolume associated with a single spatial resolution
element pixel with $\Delta z=1$ and ($L^*$, $\phi^{*}$, $\alpha$ ) are the
parameters of the LF. 

A conservative upper limit of $\mu=25$ was adopted in the vicinity of the
critical lines in order to avoid singularities in the magnification/dilution
determination. This is justified by the finite resolution of instruments, and
the limited knowledge on the precise location of the critical lines at such
high-z (typically $\sim1''$). 

When exploring the impact of cluster redshift, we assumed no evolution in the physical parameters 
$r_{core}$, $r_{cut}$ and $\sigma$ of individual potentials, thus keeping the
total mass of the cluster constant in this process.  
Variations in $z_{c}$ from the original redshift $z_{c0}$ of the cluster
produce a geometrical effect on the central positions (${x_c}_i$, ${y_c}_i$)
of each PIEMD potential $i$, measured from a reference position ($x_0$, $y_0$)
which coincides with the center of the BCG:  

\begin{eqnarray}
{x_c}_i(z_{c})-x_0= \frac{D_A(z_{c})}{D_A(z_{c_0})}({x_c}_i(z_{c_0})-x_0)\nonumber \\
{y_c}_i(z_{c})-y_0= \frac{D_A(z_{c})}{D_A(z_{c_0})}({y_c}_i(z_{c_0})-y_0)
\label{homo}
\end{eqnarray}

Similarly, we produced fiducial masks at different cluster redshifts $z_{c}$
based on the reference mask at $z_{c0}$ and adjusting the parameters of the
galaxy ellipses, applying the same scaling relations on ($x_c$,$y_c$). The
sizes $a$ and $b$ were scaled by $D_A(z_{c0})/D_A(z_{c})$.  


\section{\label{results} Results}

As discussed above, lensing introduces two opposite trends on the observed
sample as compared to blank fields : gravitational magnification by a factor
$\mu$, increasing the expected number of sources and thus the total number of
objects, and reduction of the effective surface by the same factor thus
leading to a dilution in expected counts. This effect was first studied by
\citet{Broadhurst}.  

If we consider, for a given redshift z, the cumulative abundance of sources
(per unit of solid angle) with a luminosity greater than L and by redshift
bin, the magnification bias will change depending on $\mu$ according to 

\begin{align}
n'_{lensing}(>L,z)&=N(>L/\mu,z)/\mu(z) \\ \ & \simeq \mu^{\beta(z)-1}n(>L,z)
\end{align}

where $\beta$ is the logarithmic slope of $n(L,z)$ assuming that this function
is well represented by a power law in this interval of luminosities :
$\beta=-d($ln$ n)/d($ln$ L)$. The effect on number counts is as follows: 
\begin{itemize}
 \item if $\beta(z) > 1$ the number counts will increase with respect to a blank field, and
 \item if $\beta(z) < 1$ there will be an opposite trend: i.e. a depletion in number counts
\end{itemize}
With increasing depth, the $\beta$ parameter will decrease in a greater or
lesser amount depending on the LF, the FOV (because it determines the mean
$\mu$) and the redshift of sources, leading to a depletion in number counts in
lensing fields as compared to a blank fields. With these simple considerations,
we expect lensing clusters to be more efficient than blank fields in
relatively shallow surveys. 

The efficiency of using lensing clusters as gravitational telescopes to find
high-z galaxies can be quantified with simple assumptions taking advantage of
the properties of the sources explained in Sect.~\ref{method}. In this
section, we discuss the results obtained by exploring the relevant intervals
in the parameter space. We present a comparison between the number counts
expected in lensing and blank fields, as a function of source redshift and for
different LFs. The influence of lensing cluster properties and redshift is
also studied, as well as the expected field to field variance. 

\subsection{\label{FOV} The influence of the field of view}

Here we discuss on the influence of the FOV in the simulations for typical
surveys. The influence of the limiting magnitude will be discussed in
Sect.~\ref{discussion}. 
Three different FOV are considered here: 
\begin{itemize}
\item $6'\times6'$ (``EMIR-like'' aperture)
\item $2.2'\times2.2'$ (``JWST-like'' or ``WFC3/HST'' aperture)
\item $0.85'\times0.85'$ (NICMOS/HST aperture)
\end{itemize}
In the last case, the FOV is centered along the critical lines in order to
achieve the highest mean magnification (Fig.~\ref{cluster}).  

\begin{figure}
 \centering
 \includegraphics[width=9.5cm]{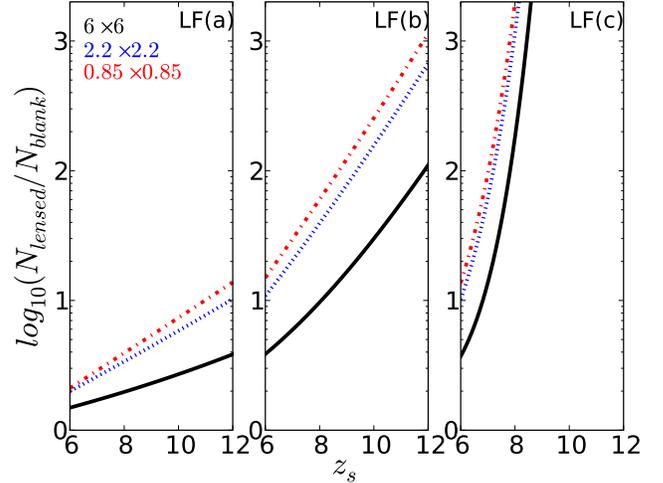}
 \caption{Relative gain in number counts between lensing and blank fields as a
   function of the source redshift, with $\Delta z=1$, for different fields of
   view: $6'\times6'$ in black, $2.2'\times2.2'$ in blue dotted line and
   $0.85'\times0.85'$ in red dot-dashed line (from bottom to top). The three
   panels from left to right represent the 3 values of the LF, (a), (b) and
   (c) respectively. \label{gain}} 
\end{figure}

The limiting magnitude is $AB\ltapprox25.5$, a value ranging between
$L_{*}$(z=6) and $3L_{*}-5L_{*}$ at redshift $z_{s}\sim7$ to $10$. The cluster
model corresponds to AC114, but the results are qualitatively the same with
other models. Fig.~\ref{gain} displays the relative gain in number counts
between lensing and blank fields as a function of sources redshift, for the
three values of the FOV mentioned above, and for the three LF adopted in the
present simulations. 

The largest gain is obtained for the smallest FOV, as expected from
geometrical considerations, because the mean magnification decreases with
increasing FOV, and in this case $\beta \gtapprox 1$ given the shallow depth
($AB\ltapprox25.5$). For a given FOV, the difference between lensing and blank
field results strongly depends on the shape of the LF. Hence, the comparison
between lensing and blank field number counts is likely to yield strong
constraints on the LF, provided that field-to-field variance is small
enough. This issue is addressed in Sect.~\ref{variance}. In the following
subsections, we adopt a $6'\times6'$ FOV unless otherwise indicated. 

\subsection{\label{lens_blank} Lensing versus Blank field efficiency}

\begin{figure*}
 \centering
 \includegraphics[width=19cm]{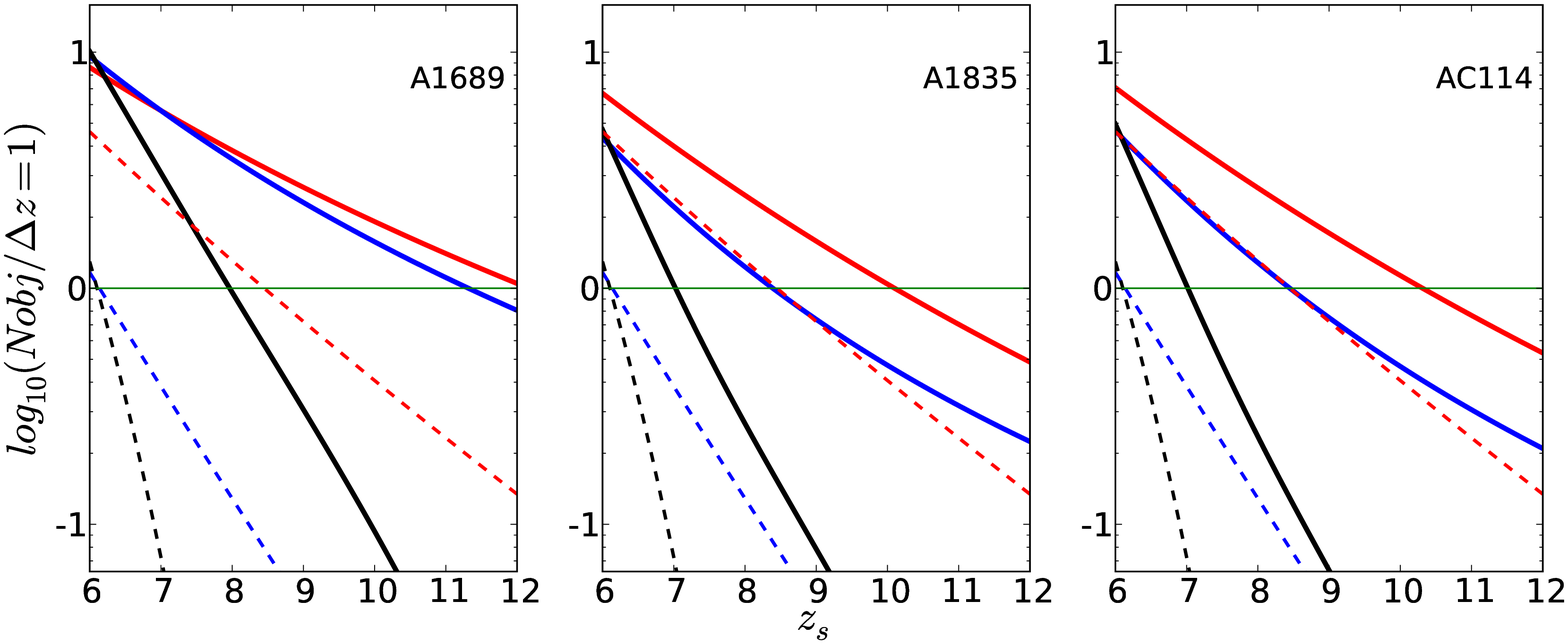}
 \caption{Comparison between the expected number counts of galaxies in a typical $6'\times6'$ FOV, up to $H_{AB} \le 25.5$, per redshift bin $\Delta z=1$, in a blank field (dashed lines) and in the field of lensing cluster (solid lines) (left to right respectively A1689, A1835 and AC114).
Expected counts are obtained by the integration of 3 different luminosity functions (a), (b) and (c) from top to bottom.
The limit of one source detected in the field of view is indicated by an horizontal line to guide the eye.\label{fig_lens_blank}}

\end{figure*}

In this section, we study the effects of lensing clusters on source counts,
using lensing models for the three reference clusters. We compute the expected
number of sources brighter than $m_0$, the typical apparent magnitude reached
in ground-based near-IR surveys. The comparison between expected number counts
of galaxies in a typical $6'\times6'$ FOV, up to $m_0\le25.5$, per redshift
bin $\Delta z=1$, in a blank field and in the field of a strong lensing
cluster are presented in Fig.~\ref{fig_lens_blank} in logarithmic scale.  

We also estimate the error on number counts due to the uncertainties on
magnification factors (Sect.~\ref{lenstool}). The choice of the LF has no
influence on the following results. Field to field variance dominate the error
budget whatever the regime. Statistical errors and systematic errors on
lensing models are smaller but not negligible as their contribution is less
sensitive than field to field variance to the number of objects. In
particular, when the number of detected sources is relatively high (i.e. when
field to field variance is relatively small), they reach $\sim15\%$ of the
error budget in the worst cases (e.g. for LF(a) at $z\sim6$ in a $6'\times6'$
FOV), and typically $\le2\%$ when the number of sources is relatively small
(e.g. for LF(c) at $z\sim8$, for any FOV). 

As shown in Fig.~\ref{fig_lens_blank}, the presence of a strong lensing
cluster has a dramatic effect on the observed number of sources, with a
positive magnification effect. 
Strong lensing fields are a factor between 2 and 10 more efficient than blank
fields for the most optimistic LF (a), the gain increasing for the LFs (b) and
(c), reaching a factor between 10 and 100 in the $z\sim6-12$ domain. A
positive magnification bias is observed, increasing with the redshift of the
sources, and also increasing from optimistic to pessimistic values of the
LF. This trend is indeed expected given the steep shape of the LF around the
typical luminosity limits achieved in ground-based ``shallow'' surveys. 

Quantitatively (cf Table~\ref{tab_lens_blank} and Fig.~\ref{fig_lens_blank}),
if the LF for LBGs was nearly constant between $z\sim4$ and 12, we could
always detect at least one object over the redshift range of interest. At
$z\sim6$, we expect up to between 7-10 sources, and at $z\sim12$ between 0.7
and 1 galaxies should be detected in a lensing field. Even in a blank field,
until $z\sim8-9$ at least one LBG could be found in such a large field of
view. With more realistic (pessimistic) values of the LF
(e.g. \citealt{Bouwens06,Bouwens08}), blank fields are particularly
inefficient as compared to lensing fields. The size of the surveyed area would
need to increase by at least a factor of $\sim10$ in order to reach a number
of detections similar to the one achieved in a lensing field around
$z\sim6-8$, and this factor increases with redshift. Note however that given a
limiting (apparent) magnitude, blank and lensing fields do not explore the
same intrinsic luminosities (see also Sect.\ref{discussion}). 

As seen in Fig.~\ref{fig_lens_blank} and Table~\ref{tab_lens_blank}, there are
also some differences between the results found in the three lensing clusters,
although they are smaller than the differences between lensing and blank
fields for a given LF. The number of expected sources behind A1689 is a factor
of two (at $z\sim6$) and a factor of three ($z\sim8$) larger than in the other
clusters for the realistic LFs (b) and (c), whereas the difference is only
$\sim10-30\%$ for LF (a). The influence of lensing properties is studied in
Sect.~\ref{clusters}. 

From the results above, it seems that lensing fields allow us to detect a
larger number of $z\gtapprox6$ sources based on their UV continuum, with some
cluster to cluster differences. This result is essentially due to the shape of
the LF. For magnitude limited samples selected within a given field of view,
the positive magnification bias increases with the redshift of the sources and
decreases with both the depth of the survey and the size of the surveyed
area. The last trend is purely geometric, as discussed in the previous
section. The differential behaviour between blank and lensing regimes strongly
depends on the shape of the LF. The comparison between blank and lensing field
observations could be of potential interest in constraining the LF, provided
that field-to-field variance is sufficiently small. This issue is addressed in
the following sections. 

\begin{table*}
\centering
\caption{Total number of objects expected within a $6'\times6'$ FOV (up to
  $H_{AB}\le 25.5$, $\Delta z=1$) from the three LF adopted in these
  simulations.
Uncertainties correspond to $1-\sigma$ level in magnification and lensing
  modeling. 
\label{tab_lens_blank}
}

\begin{tabular}{c||c c c||c c c|c c c|c c c}

\multicolumn{1}{c||}{} & \multicolumn{3}{|c||}{Blank field} & \multicolumn{9}{c}{Lensed field}\\
\cline{5-13}
\multicolumn{1}{c||}{} & \multicolumn{3}{|c||}{} & \multicolumn{3}{|c|}{A1689} & \multicolumn{3}{c|}{A1835} & \multicolumn{3}{c}{AC114} \\
\multicolumn{1}{c||}{} & \multicolumn{3}{|c||}{} & \multicolumn{3}{|c|}{$z_c=0.184$} & \multicolumn{3}{c|}{$z_c=0.253$} & \multicolumn{3}{c}{$z_c=0.310$} \\
\hline
\hline
\multicolumn{1}{c||}{z} &  \multicolumn{1}{|c}{$6$} &  \multicolumn{1}{c}{$7$} &  \multicolumn{1}{c||}{$8$} &  \multicolumn{1}{|c}{$6$} &  \multicolumn{1}{c}{$7$} &  \multicolumn{1}{c|}{$8$} &  \multicolumn{1}{|c}{$6$} &  \multicolumn{1}{c}{$7$} &  \multicolumn{1}{c|}{$8$} &  \multicolumn{1}{|c}{$6$} &  \multicolumn{1}{c}{$7$} &  \multicolumn{1}{c}{$8$} \\

LF (a) & 4.61 & 2.43 & 1.31 & $8.63^{+0.86}_{-0.76}$ & $5.66^{+0.62}_{-0.59}$ & $3.83^{+0.47}_{-0.44}$ & $6.69^{+0.61}_{-0.58}$ & $4.00^{+0.42}_{-0.39}$ & $2.48^{+0.29}_{-0.27}$ & $7.07^{+0.65}_{-0.61}$ & $4.27^{+0.44}_{-0.41}$ & $2.66^{+0.31}_{-0.29}$ \\
 & & & & & & & & & & & & \\
LF (b) & 1.30 & 0.39 & 0.13 & $10.07^{+1.808}_{-1.608}$ & $5.66^{+1.24}_{-1.08}$ & $3.53^{+0.89}_{-0.75}$ & $4.69^{+0.78}_{-0.69}$ & $2.21^{+0.47}_{-0.41}$ & $1.23^{+0.30}_{-0.25}$ & $4.96^{+0.80}_{-0.71}$ & $2.35^{+0.47}_{-0.41}$ & $1.28^{+0.30}_{-0.25}$ \\
 & & & & & & & & & & & & \\
LF (c) & 1.30 & 0.07 & $<0.01$ & $10.07^{+1.808}_{-1.608}$ & $3.08^{+0.84}_{-0.71}$ & $0.96^{+0.38}_{-0.29}$ & $4.69^{+0.78}_{-0.69}$ & $1.04^{+0.27}_{-0.23}$ & $0.28^{+0.10}_{-0.08}$ & $4.96^{+0.80}_{-0.71}$ & $1.03^{+0.26}_{-0.22}$ & $0.24^{+0.09}_{-0.07}$\\

\end{tabular}

\end{table*}

\subsection{\label{zcluster} Redshift of the lensing cluster}

\begin{figure*}
 \centering
 \includegraphics[width=19cm]{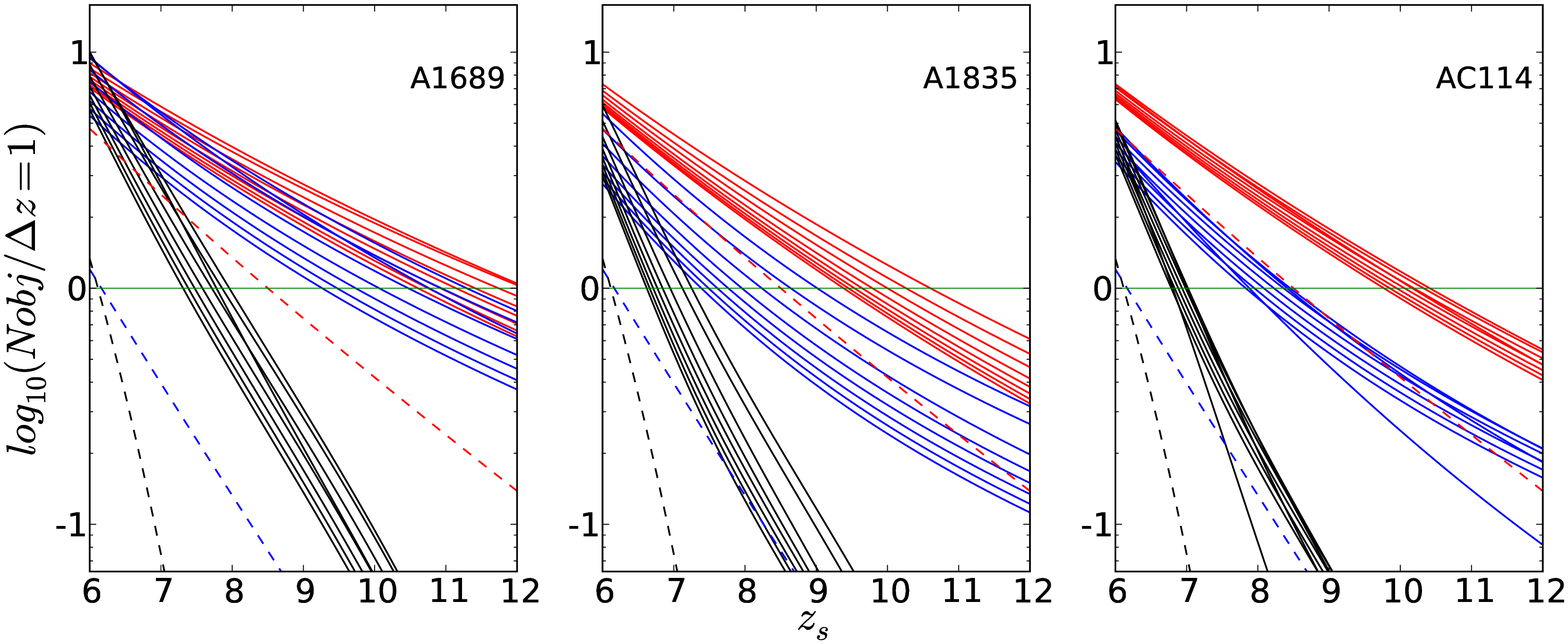}
 \caption{The same as Fig. \ref{fig_lens_blank} but using different
assumptions for the redshift of the clusters, with $z_c\ \epsilon$ [0.1, 0.8]
and a 0.1 step (for more details see 
   Sect. \ref{zcluster})\label{fig_lens_blank_zc}} 
\end{figure*} 

The redshift of the lensing cluster is a potentially important parameter when
defining an ``ideal'' sample of gravitational telescopes. Based on geometrical
considerations, we expect the magnification bias to decrease with cluster
redshift ($z_c$) after reaching a maximum efficiency at some point, depending
on cluster properties and the size of the surveyed field. The field of view
considered here is typically a few square arcminutes, essentially including
the region around the critical lines where magnification factors are the
highest. Further down, we study the impact of $z_c$ on the magnification
bias. 
  
\begin{figure}[!h]
 \centering
 \includegraphics[width=9.5cm]{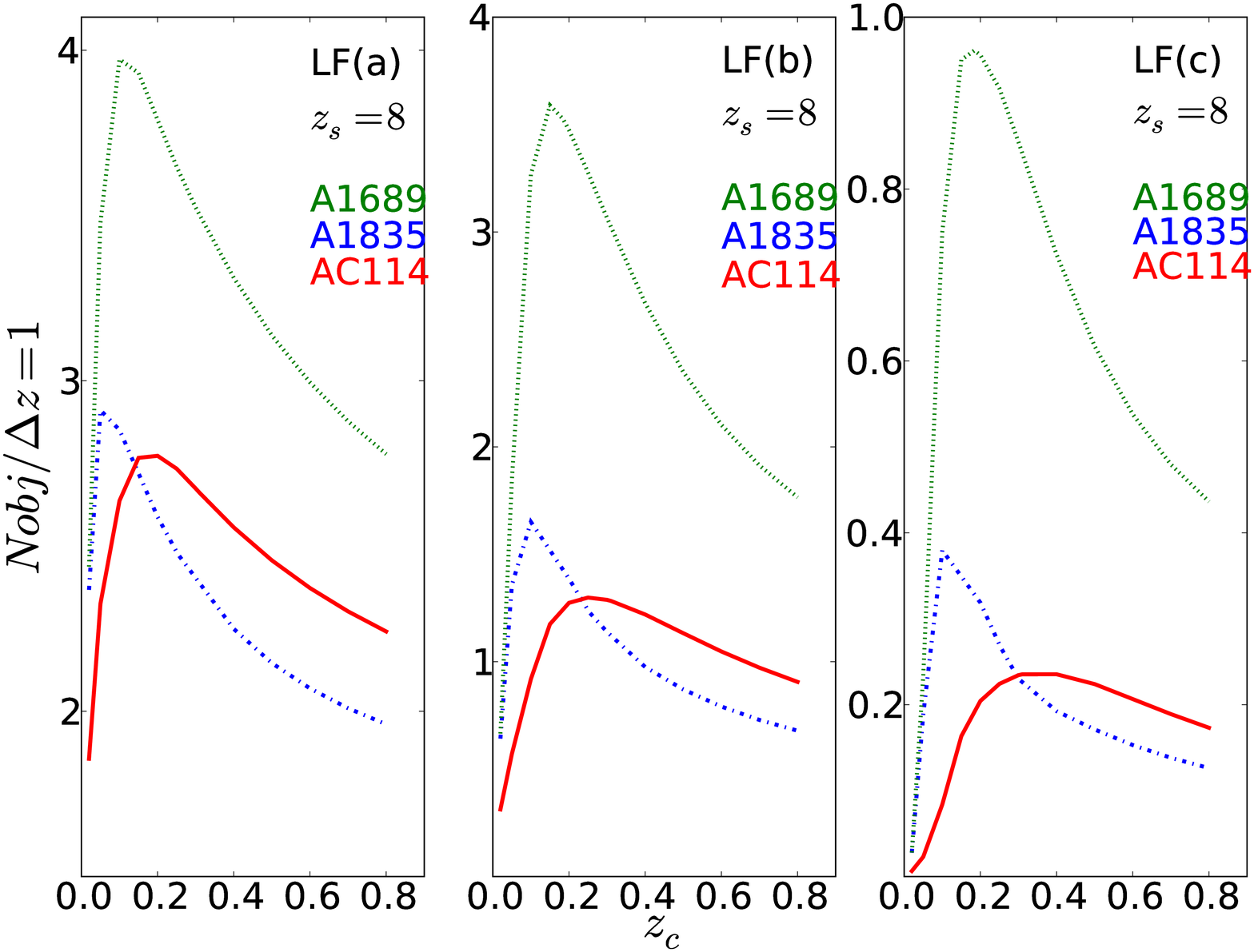}
 \caption{Expected number of objects as a function of the cluster redshift for
   a fixed redshift of sources ($z_s=8$) and with the same depth and field
   size as in previous section. Three cluster models are displayed: A1689
   (green dotted line), A1835 (blue dot-dashed line) and AC114 (red solid
   line). Panels from left to right display respectively LF (a), (b) and
   (c).\label{fig_zc}} 
\end{figure}

Using the non-evolution assumption presented in Sect.~\ref{implementation}, we
compute the expected number counts for the three reference models (A1689,
A1835 and AC114) with cluster redshifts ranging between z=0.1 and 0.8, with a
$\Delta z=0.1$ step. A step $\Delta z=0.05$ is used in the z=0.1-0.3 interval,
in order to refine the sampling around the maximum value. We use the same
depth and field size as in previous section. The effect of cluster redshift is
clearily seen in Fig.~\ref{fig_zc} representing the number of objects as a
function of cluster redshift (for the three reference models), at a fixed
redshift of z=8 for sources. 

The global effect of $z_c$ on number counts as a function of the source
redshift is displayed in Fig.~\ref{fig_lens_blank_zc}. This figure directly
compares to Fig.~\ref{fig_lens_blank} in the previous
section. Table~\ref{tab_max_zc} presents the $z_c$ value which corresponds to
a maximum in the expected number counts at z=8. This value depends slightly on
the source redshift and LF. In addition to the $\Delta z_c=+-0.05$ when
changing the LF, there is also an increase of $z_c$ with higher values of
$z_s$, up to +0.05 towards $z_s=12$. The search efficiency of distant galaxies
in lensing fields is maximised when using clusters at low redshift
($z_c\sim0.1-0.3$). Although the field of view considered here is relatively
large for near-IR surveys and close to present-day cameras, it is the limiting
factor at $z_c\ltapprox0.1$, where an increasing fraction of the
strong-magnification area is lost with decreasing $z_c$. Also, in this $z_c$
regime, the field of view concentrates on the central region of the cluster
where bright cluster galaxies mask an important fraction of the
strong-magnification area. The high magnification region represents an
increasingly small percentage of the field with increasing $z_c$. Number
counts in this regime asymptotically tend towards a limiting value with
increasing $z_c$ (Fig.~\ref{fig_lens_blank_zc}), wich still is a substantial
gain with respect to a blank field of the same size. The non-evolution
assumption in cluster properties has a weak effect on this conclusion. Indeed,
clusters far from relaxation will be even more ineffective as a gravitational
lenses in the strongest magnification regime. The results obtained are hence
an optimistic upper limit on number counts for realistic clusters beyond the
optimal regime $z_c\sim0.1-0.3$. 

\begin{table}[!h]
\centering 
\caption{Redshift of the cluster which maximizes the number of objects
  detected at z=8 for the three LF respectively from top to bottom (a), (b)
  and (c)\label{tab_max_zc}} 

\begin{tabular}{c|c c c} 
LF & A1689 & A1835 & AC114 \\
\hline
actual $z_c$ & 0.184 & 0.253 & 0.310 \\
(a) & 0.10 & 0.05 & 0.25 \\
(b) & 0.15 & 0.10 & 0.25 \\
(c) & 0.20 & 0.10 & 0.30 \\

\end{tabular}
\end{table}

\subsection{\label{clusters} Influence of lensing cluster properties}

\begin{figure*}
\centering 

\begin{tabular}{c c}
\includegraphics[width=8cm]{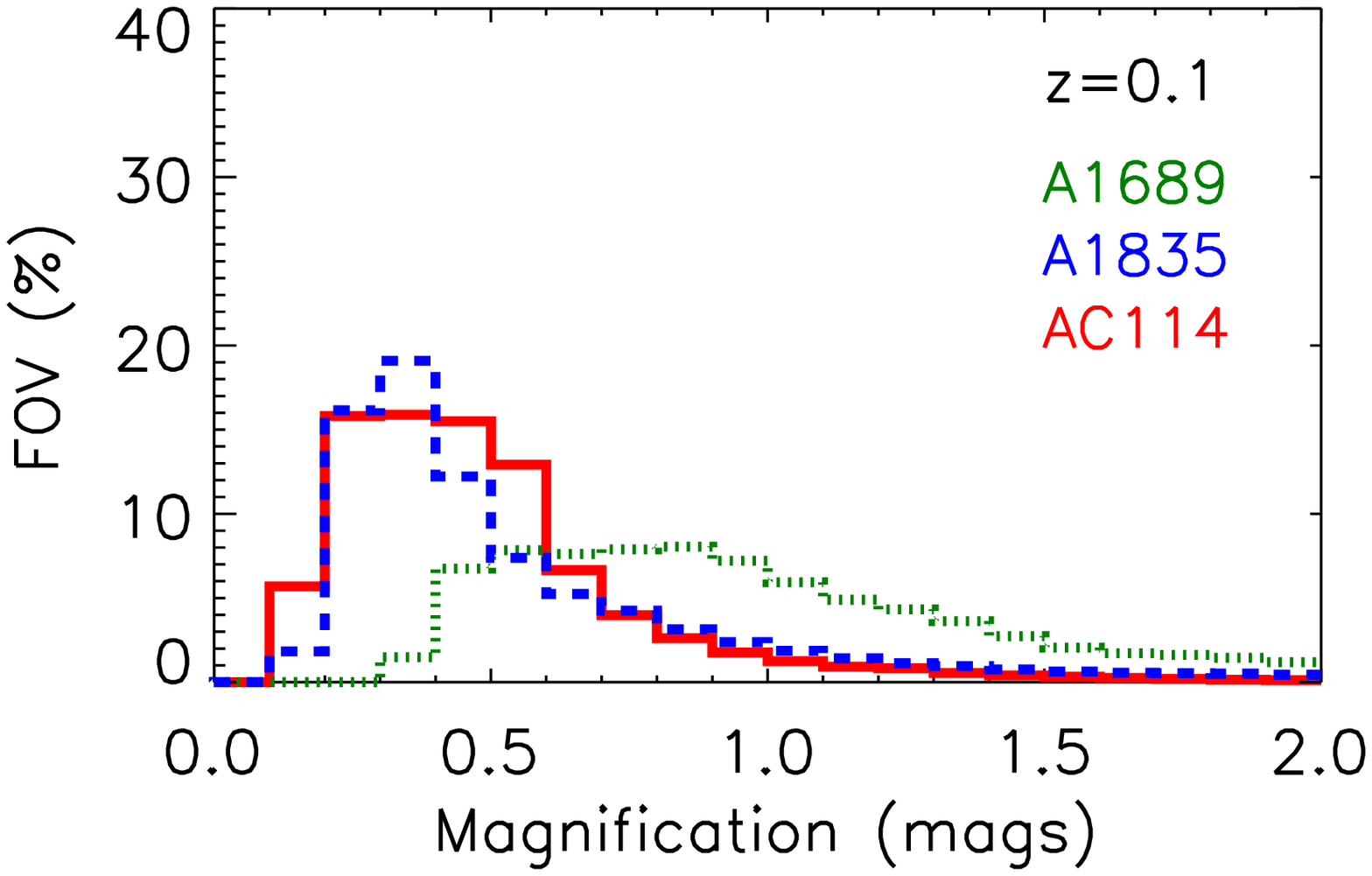} 
\includegraphics[width=8cm]{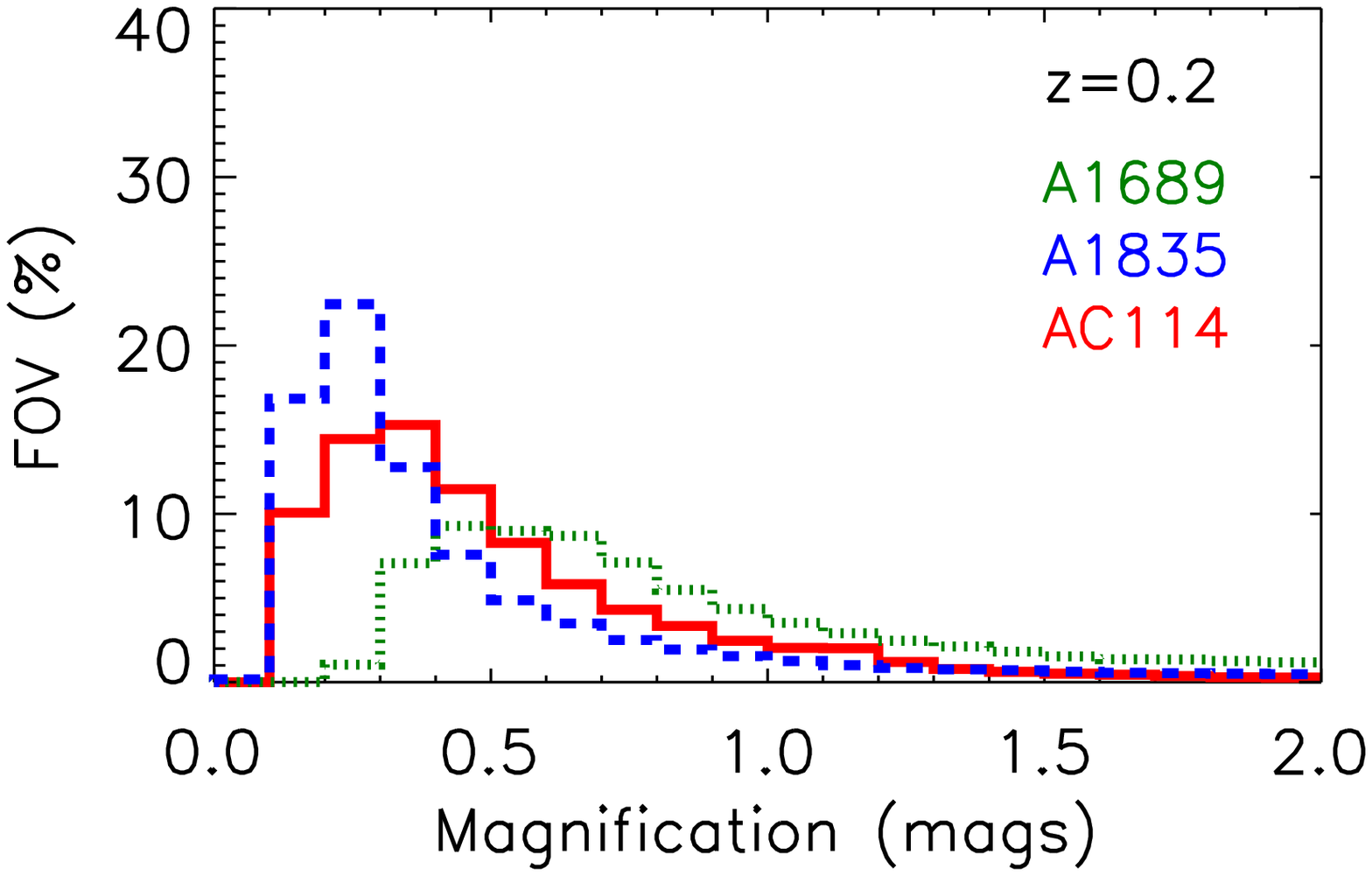} \\
\includegraphics[width=8cm]{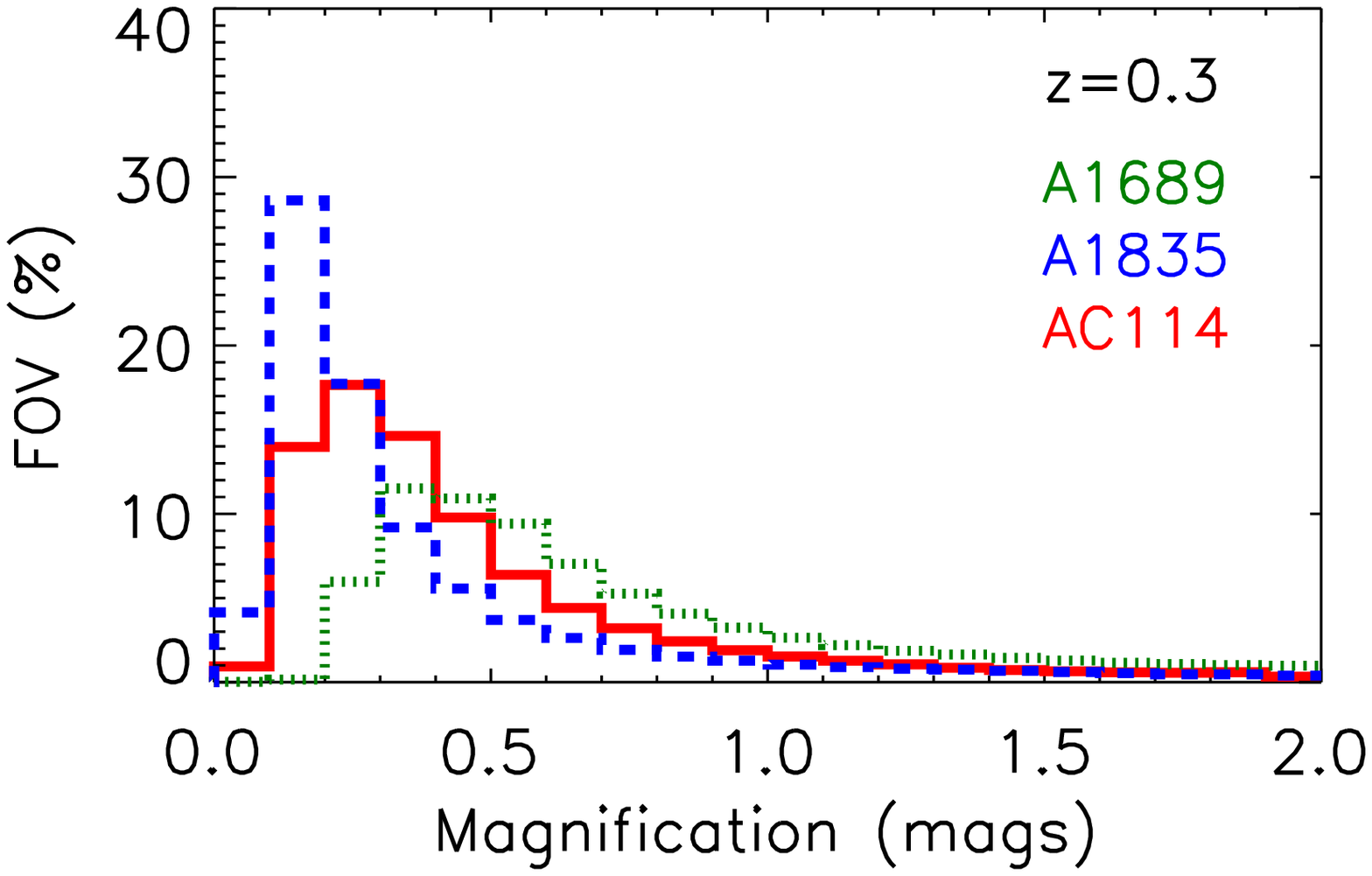} 
\includegraphics[width=8cm]{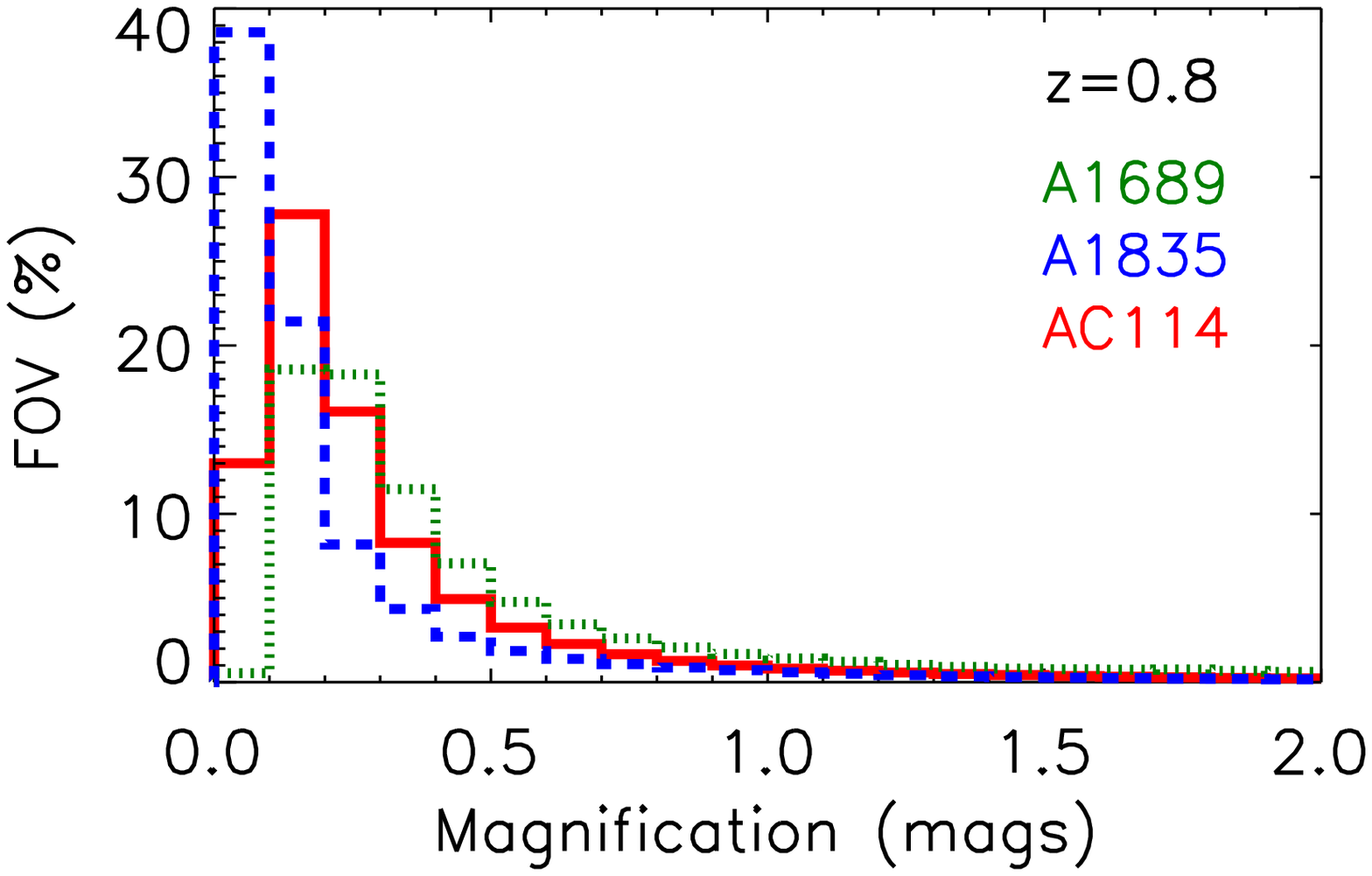} \\
\end{tabular}

\caption{Histogram representing the percentage of the surface ($6'\times6'$
  FOV) as a function of the magnification for different redshifts of cluster
  ($z_c$=0.1, 0.2, 0.3, 0.8), using the same color codes for the three
  clusters as in Fig.~\ref{fig_zc} (A1689: green dotted line, A1835: blue
  dot-dashed line and AC114: red solid line) \label{fig_histo_mu}} 

\end{figure*}

In this section we focus on the differences between lensing cluster properties
and their influence on expected source counts. As seen in previous sections,
A1689-like clusters are expected to be more efficient irrespective of the
cluster redshift. To understand this effect, we study the magnification
regimes for a reference source plane fixed at $z_s=8$. 
The distribution of the magnification regimes in the image plane varies from
cluster to cluster. Histograms in Fig. \ref{fig_histo_mu} represent the
percentage of the image plane (for the $6'\times6'$ FOV) as a function of the
associated magnification. To perform this calculation, cluster redshifts were
standardized to identical values for a better understanding of the
phenomenom. As seen in the figure, A1689 shows a different regime at
$z_c\ltapprox0.8$ as compared to the other clusters. While the percentage of
the surface affected by strong magnification ($\mu>10$) does not exceed $5\%$
in A1835 and AC114, it is as high as $8-10\%$ in A1689, depending on
$z_c$. Nevertheless, this difference between clusters tends to fade with
increasing cluster redshift due to projection effects, the fraction of highly
magnified pixels becoming smaller with respect to the whole FOV. We also note
that AC114 and A1835 models have a similar behaviour with minor differences
(A1835-like clusters being more effective at very low $z_c$ while the AC114
model is more efficient for intermediate $z_c$).  

\begin{figure}
 \centering
 \includegraphics[width=9.0cm]{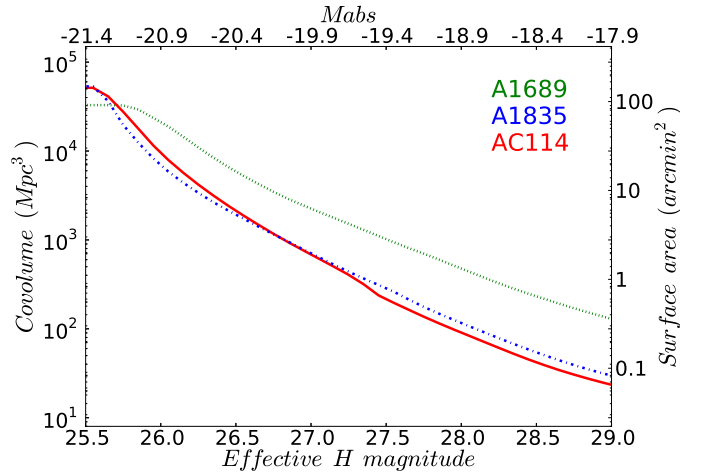}
 \caption{The effective (lensing corrected) covolume sampled at z=6.5-7.5 by
   each cluster is given as a function of effective $H_{AB}$ magnitude limit,
   for a magnitude-limited survey of $6'\times6'$ FOV with $H_{AB}\le
   25.5$. The three clusters are displayed with the same colors and line
   codes as in Fig.~\ref{fig_zc}.\label{fig_cv}} 
\end{figure}

Another way of understanding this phenomenom is presented in the
Fig.~\ref{fig_cv}, where the effective covolume for the $6'\times6'$ FOV is
traced as a function of the effective magnitude, for a magnitude-limited
survey with $AB\leq25.5$. Magnification in lensing fields provides an enhanced
depth for a magnitude-limited survey, where the effective (lensing corrected)
covolume surveyed decreases with increasing effective depth. The behavior of
A1689 in Fig.~\ref{fig_cv} illustrates the situation for this particularly
efficient cluster, allowing us to study a $\sim100 Mpc^3$ volume to
$AB\sim29.0$ with a relatively modest observational investment. Except for
some particularly efficient lensing clusters (such as A1689), most lensing
fields should behave the same way as A1835 or AC114. 

\subsection{\label{variance} Field to field variance}

In this section, we address the expected field-to-field variance affecting our
previous results in order to estimate its impact in blank and lensed
fields. We used two different approaches: the two-point correlation function
estimation proposed by \citet{Trenti} and a pencil beam tracer through the
Millenium simulation. 

The first estimate is based on the method implemented by \citet{Trenti}. This
method for the calculation of the cosmic variance is based on the two points
correlation function $\xi(r)$ of the sample (\citealt{Peebles}). Field o field
variance is given by  
\begin{equation}
 \sigma_{v}^2=\frac{\int_{V}\int_{V} \mathrm d^3x_{1} \mathrm
   d^3x_{2}\xi(|\vec{x_{1}}-\vec{x_{2}}|)}{\int_{V}\int_{V} \mathrm d^3x_{1}
   \mathrm d^3x_{2}} 
\end{equation}
where V represents the volume of the survey.

We focus on the redshift interval $z\sim6-8$ using the present ``shallow''
survey parameters (see Sect.~\ref{efficiency}), both in blank and lensing
fields. Here we use the same parameters as in Sect.~\ref{lens_blank},
i.e. typical FOV $6'\times6'$, $m_{AB}=25.5$ and $\Delta_z=1$. 

We define the total fractional error of the counts N following \citet{Trenti}
(this is the so-called field-to-field standard deviation or the, again,
improperly called "cosmic variance") as :  
\begin{equation}
 v_{r}=\frac{\sqrt{\langle N^2 \rangle-\langle N \rangle ^2}}{\langle N \rangle}
\end{equation}

Results are presented in the Table \ref{tab_vr} for the three LF considered,
and for the three typical clusters used in the simulations. We note an
important field to field variance with such a limiting magnitude
($AB\sim25.5$) either in blank or lensing fields due to the small number
counts previously derived from calculations (see Table
\ref{tab_lens_blank}). Nevertheless, the variance is smaller behind
gravitational telescopes with the same differential trends mentioned before
between the three clusters, i.e. A1689 exhibits a stronger magnification bias
(see Sect.~\ref{clusters}) than the other clusters which have a similar
behavior. Besides, with increasing redshift of sources, the expected number
counts decrease leading to a larger field to field variance. 

The second estimate was based on the Millennium simulation, carried out by the
Virgo Consortium and described in detail in \cite{Springel}
and \cite{Lemson06}. 
The simulation
follows N = $2160^{3}$ particles of mass $8.6 \ 10^8 \ h^{-1} \ \msun$ within
a co-moving box of size $500 \ h^{-1} \ Mpc$ on a side. The cosmological model
is a $\Lambda$CDM model with small differences in the cosmological parameters
adopted in Sect. \ref{intro}, but without impact on the final results. These
cosmological parameters are consistent with recent determinations from the
combined analysis of the 2dFGRS and 3rd year WMAP data
(\citealt{Sanchez}). Given its high resolution and large volume, the
Millennium simulation allows us to follow in enough details the formation
history of a representative sample of high redshift galaxy environments. With
these prescriptions and a realistic beam tracer we can study the
field-to-field variations in the number counts of star forming galaxies at the
epoch of interest. 

Our pencil beam tracer is similar to the one developed by
\citet{Kitzbichler}. We trace through the simulation box a parallelepiped
where the base is a parallelogram, whose size is given by the reference field
of view in comoving units,  
and the depth is the comoving depth arbitrarily taken to $\Delta z = 1$. The
variation of angular distance versus redshift in the redshift interval of the
selection window considered was properly taken into account. This $500
h^{-1}Mpc$ edge box is more than $2000$ times larger than the effective volume
probed by the $6'\times6'$ FOV : $\sim 10.7\times10.7\times323.0 (Mpc/h)^3$
for instance at z=6. We carried out $10000$ Monte Carlo realizations of the
beam-tracing procedure by randomly varying the initial position of the beam in
order to calculate the typical number counts of galaxies and the associated
standard deviation in the field of view with the same hypothesis.  

Although this procedure is well suited to determine the field to field
variance, several studies on this topic suggest an overprediction on the
abundance of massive galaxies at high redshift (e.g. \citealt{White}). For
this reason, we consider this second approach as a cross check yielding a
lower limit for the field to field variance. Results obtained from the
Millenium simulation are displayed in Table \ref{tab_vr_millenium}. They are
in fair agreement with those obtained with the first method. 

Field to field variance on number counts obviously depends on the depth of the
survey. In order to compare our results with existing photometric surveys, we
calculated the number counts of sources in blank and lensing fields (here
AC114) with the evolving LF(c) for different deeper magnitude limits
($m_{lim}=27.0$, $28.0$ and $29.0$), in our reference field of view using the
same parameters as in Sect.~\ref{lens_blank} (see
Table~\ref{tab_vr_allmag}). The correlation function was used to derive the
cosmic variance. The total fractional error ($v_r$) strongly decreases with
increasing photometric depth, as expected given the increasing number of
sources detected in such a large FOV (e.g. at $m_{lim}=29.0$, the total number
of sources is ~1000 times larger than in the ``shallow'' survey), both in
blank and lensing fields. The fractional error appears slightly larger in
lensing than in blank fields at z=6, but this effect reverses with increasing
source redshift.  
These estimates for the blank field can be compared to present-day
surveys. For instance, the field to field variations obtained by
\citet{Bouwens06} for a single ACS pointing at $z\sim6$ for a limiting
magnitude $z_{850}\sim29$ is 35\%. Using the same observational constraints
(FOV, depth, ...), our simulations yield a $v_r\sim30\%$, a value which is
smaller but fairly compatible with the results quoted by \citet{Bouwens06}. 

\begin{table*}
\centering

\caption{Number counts, field to field variance calculated with the
  correlation function both in
  blank and lensing fields, for z = 6, 7 and 8 within a $6'\times6'$ FOV, for
  a shallow survey with $AB\le25.5$.
Field to field variance in lensing fields includes $1-\sigma$ magnification errors.
\label{tab_vr}} 
\begin{tabular}{c||c c c||c c c|c c c|c c c}

\multicolumn{1}{c||}{} & \multicolumn{3}{|c||}{Blank field} & \multicolumn{9}{c}{Lensed field}\\
\cline{5-13}
\multicolumn{1}{c||}{} & \multicolumn{3}{|c||}{} & \multicolumn{3}{|c|}{A1689} & \multicolumn{3}{c|}{A1835} & \multicolumn{3}{c}{AC114} \\
\hline
\hline
\multicolumn{1}{c||}{$z_s$} &  \multicolumn{1}{|c}{$6$} &  \multicolumn{1}{c}{$7$} &  \multicolumn{1}{c||}{$8$} &  \multicolumn{1}{|c}{$6$} &  \multicolumn{1}{c}{$7$} &  \multicolumn{1}{c|}{$8$} &  \multicolumn{1}{|c}{$6$} &  \multicolumn{1}{c}{$7$} &  \multicolumn{1}{c|}{$8$} &  \multicolumn{1}{|c}{$6$} &  \multicolumn{1}{c}{$7$} &  \multicolumn{1}{c}{$8$} \\

LF (a)& 4.61 & 2.43 & 1.31 & $8.63^{+0.86}_{-0.76}$ & $5.66^{+0.62}_{-0.59}$ & $3.83^{+0.47}_{-0.44}$ & $6.69^{+0.61}_{-0.58}$ & $4.00^{+0.42}_{-0.39}$ & $2.48^{+0.29}_{-0.27}$ & $7.07^{+0.65}_{-0.61}$ & $4.27^{+0.44}_{-0.41}$ & $2.66^{+0.31}_{-0.29}$ \\
& & & & & & & & & & & & \\
$v_{r}$                    & 58\%  & 75\% & 97\% & $47^{+1}_{-1}$\%  & $56^{+2}_{-2}$\%  & $66^{+3}_{-3}$\%  & $51^{+1}_{-1}$\%  & $63^{+2}_{-2}$\%  & $77^{+3}_{-3}$\% & $50^{+1}_{-1}$\%  & $62^{+2}_{-2}$\%  & $75^{+3}_{-3}$\% \\
 & & & & & & & & & & & & \\
 & & & & & & & & & & & & \\
LF (b)& 1.30 & 0.39 & 0.13 & $10.07^{+1.808}_{-1.608}$ & $5.66^{+1.24}_{-1.08}$ & $3.53^{+0.89}_{-0.75}$ & $4.69^{+0.78}_{-0.69}$ & $2.21^{+0.47}_{-0.41}$ & $1.23^{+0.30}_{-0.25}$ & $4.96^{+0.80}_{-0.71}$ & $2.35^{+0.47}_{-0.41}$ & $1.28^{+0.30}_{-0.25}$ \\
 & & & & & & & & & & & & \\
$v_{r}$                    & 96\% & 165\% & 283\%&  $45^{+2}_{-2}$\% & $56^{+4}_{-4}$\%  & $68^{+5}_{-6}$\%  & $60^{+3}_{-3}$\%  & $79^{+6}_{-7}$\%  & $101^{+9}_{-10}$\%& $57^{+4}_{-4}$\%  & $77^{+6}_{-6}$\%  & $100^{+8}_{-9}$\% \\
 & & & & & & & & & & & & \\
 & & & & & & & & & & & & \\
LF (c)& 1.30 & 0.07 & $<0.01$ & $10.07^{+1.808}_{-1.608}$ & $3.08^{+0.84}_{-0.71}$ & $0.96^{+0.38}_{-0.29}$ & $4.69^{+0.78}_{-0.69}$ & $1.04^{+0.27}_{-0.23}$ & $0.28^{+0.10}_{-0.08}$ & $4.96^{+0.80}_{-0.71}$ & $1.03^{+0.26}_{-0.22}$ & $0.24^{+0.09}_{-0.07}$\\
 & & & & & & & & & & & & \\
$v_{r}$                    & 96\% & 383\% & \_ & $45^{+2}_{-2}$\% & $70^{+6}_{-7}$\% & $113^{+17}_{-16}$\% & $60^{+3}_{-3}$\% & $108^{+12}_{-10}$\% & $197^{+39}_{-25}$\% & $57^{+4}_{-4}$\% & $108^{+12}_{-10}$\% & $211^{+38}_{-29}$\% \\

\end{tabular}

\end{table*}

\begin{table}
\centering 
\caption{Number counts for $m_{AB}\le25.5$ and field to field uncertainties
  ($v_r$) calculated from the Millenium simulation in a $6'\times6'$ blank
  field, for different source redshifts\label{tab_vr_millenium}} 

\begin{tabular}{c|c c c} 
      & $z_s=6$ & $z_s=7$ & $z_s=8$ \\
\hline
$\langle N \rangle$ & 2.74 & 1.10 & 0.39 \\
$v_{r}$ & 68\% & 100\% & 168\% \\

\end{tabular}
\end{table}

\begin{table*}
\centering 
\caption{Field to field variance for 3 different magnitude limits :
  $m_{lim}=27.0$, $28.0$ and $29.0$, in a $6'\times6'$ blank field and lensing
  field (behind AC114) for the LF(c).
Field to field variance in lensing fields includes $1-\sigma$ magnification errors.
\label{tab_vr_allmag}}

\begin{tabular}{c|c c|c c|c c} 
      & $m_{lim}=27.0$ &  & $m_{lim}=28.0$ &  & $m_{lim}=29.0$ &  \\
      & $\langle N \rangle$ & $v_{r}$ & $\langle N \rangle$ & $v_{r}$ & $\langle N \rangle$ & $v_{r}$ \\
\hline
Blank field \\
$z=6$ & 65.58 & 27\% & 256.71 & 21\% & 681.92 & 18\% \\
 & & & & & & \\
$z=7$ & 19.01 & 38\% & 107.27 & 26\% & 352.52 & 21\% \\
& & & & & & \\
$z=8$ & 4.17 & 62\% & 39.78 & 34\% & 170.04 & 26\% \\
& & & & & & \\
$z=10$ & 0.05 & 471\% & 2.92 & 72\% & 29.73 & 39\% \\
& & & & & & \\
$z=12$ & $<0.00$ & \_ & 0.04 & 494\% & 2.56 & 77\% \\

\hline
Lensing field \\
$z=6$ & $83.88^{+6.60}_{-6.28}$ & $26^{+0}_{-0}$\% & $253.92^{+13.20}_{-12.86}$ & $23^{+0}_{-0}$\% & $537.27^{+17.23}_{-17.22}$ & $22^{+0}_{-0}$\% \\
& & & & & & \\
$z=7$ & $31.05^{+3.21}_{-3.00}$ & $33^{+1}_{-1}$\% & $122.49^{+8.16}_{-7.85}$ & $26^{+1}_{-1}$\% & $318.64^{+13.88}_{-13.66}$ & $24^{+0}_{-0}$\% \\
& & & & & & \\
$z=8$ & $10.03^{+1.39}_{-1.27}$ & $45^{+1}_{-2}$\% & $54.01^{+4.57}_{-4.33}$ & $31^{+1}_{-1}$\% & $174.64^{+9.75}_{-9.46}$ & $26^{+0}_{-0}$\% \\
& & & & & & \\
$z=10$ & $0.72^{+0.20}_{-0.16}$ & $129^{+14}_{-14}$\% & $7.44^{+1.06}_{-0.96}$ & $55^{+2}_{-2}$\% & $41.28^{+3.58}_{-3.39}$ & $37^{+1}_{-2}$\% \\
& & & & & & \\
$z=12$ & $0.06^{+0.03}_{-0.02}$ & $411^{+107}_{-78}$\% & $0.64^{+0.17}_{-0.14}$ & $137^{+15}_{-14}$\% & $6.46^{+0.92}_{-0.83}$ & $59^{+2}_{-2}$\% \\
\end{tabular}
\end{table*}


\section{\label{discussion} Discussion}

\subsection{\label{efficiency} Survey parameters and efficiency}

As discussed in Sect.~\ref{simulation}, the FOV and the limiting magnitude are
two important survey parameters used in these simulations. The influence of
the FOV for a fixed limiting magnitude strongly depends on the shape of the
LF. The highest ratio in number counts between lensing and blank fields can be
achieved with the smallest FOV due to simple geometrical considerations. This
section specifically addresses the evolution on the survey efficiency in
lensing and blank fields as a function of the limiting magnitude. 

For these purposes, we use the same approach as in Sect.~\ref{lens_blank} to
derive number counts within a $6'\times6'$ FOV in blank fields and behind
lensing clusters. AC114 is used here as a representative lensing
cluster. Fig.~\ref{EMIR_mags} displays the expected number counts as a
function of the redshift of sources, for different depths ($AB\le26.0, 27.0,
28.0$ and $29.0$). An opposite trend between blank and lensing fields appears,
depending once again on the LF and on the redshift of sources. With increasing
limiting magnitude, the efficiency of the survey towards a foreground cluster
diminishes and becomes less effective than in blank fields leading to a
negative magnification bias for the faintest limiting magnitudes (e.g. for
LF(a) between $AB\sim27.0-28.0$, for LF(b) between $AB\sim28.0-29.0$ and for
LF(c) beyond $AB\sim29.0$). This trend, however, is highly sensitive to the
FOV. In particular, the negative magnification bias appears towards the
typical magnitudes achieved by space facilities (JWST). Fig.~\ref{JWST_mags}
displays the same results as in Fig.~\ref{EMIR_mags} but for a
$2.2'\times2.2'$ FOV (JWST-like). The main characteristics remain broadly
unchanged, the general trends are just exacerbated, the inversion happening to
lesser depth. 

Lensing and blank field surveys do not explore the same intrinsic
luminosities, as shown in Fig.~\ref{intrinsic_true} and~\ref{intrinsic}. These
figures compare the expected number density of sources as a function of their
intrinsic UV luminosity (or equivalent SFR) for different limiting magnitudes
ranging from $AB\ltapprox25.5$ to $29.0$. In the case of lensing fields, two
different results are given, depending on the FOV around the cluster
center. In this particular case, the source redshift is arbitrarily fixed to
$z_s=8$, assuming a strongly evolving LF(c), and the lensing cluster is
AC114. 

In summary, the number of z$>$8 sources expected at the typical depth of JWST
($AB\sim28-29$) is much higher in lensing than in blank fields if the UV LF is
rapidly evolving with redshift (LF(c)), as suggested by \citet{Bouwens08}. The
trend should be the opposite if the LF remains unchanged between $z\sim6$ and
$8$. Lensing clusters are the only way to study the faintest building blocks
of galaxies, with typical $SFR\ltapprox0.1$ to $1 \msun / yr$. On the
contrary, wide field surveys covering $10^3$ to $10^4$ $arcmin^{-2}$ are
needed to set reliable constraints on the brightest part of the LF at
$z\gtapprox6$, i.e. for galaxies with SFR$\gtapprox10 \msun / yr$. 

\subsection{\label{size} Influence of galaxy morphology and image sampling}

Gravitational magnification (e.g. in the tangential direction)
induces an elongation of images along the shear direction while preserving the
resolution in the perpendicular direction and the surface brightness of high
redshift galaxies. All the comparisons between lensing and blank
fields in our simulations assumed that observations were conducted with the
same intrument setup in terms of FOV and spatial sampling, and with the same
observational conditions, in particular the same limiting surface
brightness and PSF. However, when comparing magnitude-limited samples in lensing and
blank fields, it is worth discussing the influence of galaxy morphology and
image sampling on the present results. In particular, the evolution in the surface 
brightness of high redshift sources is susceptible to hinder the search efficiency 
in clusters if, for instance, number counts in clusters were dominated by sources below the 
limiting surface brightness.

As explained in Sect.~\ref{sources}, all the previous results have been
obtained assuming that galaxies at $z>7$ are compact as compared to spatial
sampling. Indeed, high redshift sources are expected to be very small,
typically $\ltapprox 0.``10$ on the sky, based on cosmological simulations
($e.g.$ \citealt{Barkana00}), in such a way that the high resolution
capability of JWST is needed for resolving such faint galaxies. 
Recent observations of LBGs candidates in the HUDF fully support this idea
(\citealt{Bouwens08,Bouwens09b,Oesch09}). In a recent paper, \cite{Oesch09}
measured the average intrinsic size of $z\sim 7-8$ LBGs to be $0.7 \pm0.3$
kpc. These galaxies are found to be extremely compact, with very little
evolution in their half-light radii between $z\sim6$ and 7, roughly
consistent with galaxies having constant comoving sizes, at least within the
observed luminosity domain $\sim 0.1-1 L^*(z=3)$. Smaller physical sizes are
expected for higher redshift and/or intrisically fainter galaxies, based on the
scaling of the dark matter halo mass or the disk circular velocity
(\citealt{Mo98}). This differential trend is actually observed between the
bright ($\sim 0.3-1 L^*$) and the faint ($\sim 0.12-0.3L^*$) samples of
\cite{Oesch09}.   

If all high-z galaxies exhibit the same compact and uniform morphology, the
effective mean surface brightness of a lensed galaxy will be brighter or
fainter with respect to a blank field galaxy with the same apparent magnitude
depending on the spatial resolution (in practice, the instrumental PSF).  
The majority of lensed sources should remain spatially
unresolved on their width on seeing-limited ground-based surveys, 
and even on their tangential direction up to a gravitational magnification
$\mu \sim 5-10$. Hence, the apparent surface 
brightness of a lensed source is actually brighter than that of a blank field 
galaxy of similar apparent magnitude (by roughly $-2.5 log(\mu)$ mags
for a spatially unresolved galaxy). This situation is typically found in the
``shallow and wide'' near-IR surveys discussed above (e.g. for the
$6'\times6'$ FOV), where lensing clusters are particularly efficient. 

On the contrary, for a fixed apparent magnitude, the effective mean surface
brightness of a lensed galaxy is expected to become fainter with respect to a
blank field galaxy when the image resolution is similar or better than its
(lensed maximum) half-light radius, reaching $\sim 2.5 log(\mu)$ mags in the worst
case. This situation is typically expected in the ``deep and narrow'' near-IR
surveys with space facilities. In practice, the best spatial resolution
presently achieved with HST/WFC3 in the near-IR is $\sim 0.``10$, reaching
$0.``065$ with JWST/NIRCam, i.e. the typical size of the brightest $z\sim7-8$
LBGs candidates presently identified. Therefore, the majority of lensed sources should
remain spatially unresolved on their width. A lensed source entering the
apparent-magnitude limited sample because of its magnification $\mu$ has also 
a smaller physical size, by a factor of $\mu$ (assuming a constant M/L scaling with the
halo mass) or $\mu^{1/2}$ (assuming a constant M/L scaling with the halo
circular velocity), leading to an apparent increase on its
surface brightness with respect to blank-field observations of the same
galaxy. Given the spatial resolution achieved with HST and JWST, this
intrinsic-size effect tends to compensate the image dilution described above,
in such a way that the actual surface brightness of the lensed galaxy should
get close to the surface brightness of a blank field galaxy of similar
apparent magnitude. 

   For the reasons explained above, and to the best of present knowledge, we do
not expect the apparent-magnitude limited number counts derived in clusters to
be strongly biased by sources below the limiting surface brightness, provided
that the usual scalings apply to the size of high-z sources.

\subsection{\label{example} Comparison with current survey results}

We have compared our simulation results to recent observations looking for
high-z LBGs. For instance, the discovery of a bright $z=7.6\pm0.4$ lensed
galaxy by \cite{Bradley}, with AB=24.7 (intrinsic $AB\sim-22.6$), in a
$2.5'\times2.5'$ FOV survey around A1689 is in fair agreement with our
expectations. Indeed, given the survey characteristics and including
$\sim100$\% variance for $z_s=7.6\pm0.4$, we expect between 0.2 and 0.8 
such bright objects in this lensing field, if the LF remains constant between
$z\sim4$ and $8$ (LF(a)). In case of a strongly evolving LF(c), the expected
number of sources in this survey is 0.12 (i.e. ranging between 0 and 0.5 with
$\sim200$\% variance) making the discovery of this bright source particularly
fortunate. Our results for lensing fields are also consistent with the number
of $z\sim7.5$ LBGs found by \cite{Richard08}, to the depth of their survey,
using LF(b) or (c). Quantitatively, \cite{Richard08} detected 5 sources with
12 pointings over 6 clusters. With our simulations, $2.6^{+0.7}_{-0.6}$
objects with a variance of $\sim75\%$ are expected with the LF(c) model. We
also compared with the surface density of $z\sim7$ candidates in the deep
near-IR data behind clusters obtained by \cite{Bouwens09}. \cite{Bouwens09}
found a surface density of $0.05^{+0.11}_{-0.04}$ $arcmin^{-2}$ with
$AB\ltapprox25.5$ with a typical NICMOS3 FOV. With the strongly evolving LF(c)
and the same survey characteristics used in our simulations, we expect a
surface density of 0.01 $arcmin^{-2}$ behind a typical cluster such as AC114,
with a variance of $\sim110\%$. This result shows a relatively good agreement
by taking into account field to field variance.

\begin{figure}
 \centering
 \includegraphics[width=9.5cm]{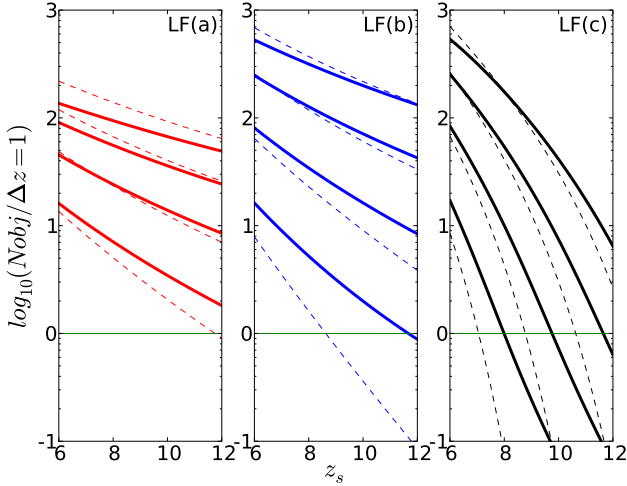}
 \caption{Expected number counts of objects as a function of the redshift of
   sources in a $6'\times6'$ FOV, for different limiting magnitudes 26.0,
   27.0, 28.0 and 29.0 from bottom to top respectively. This calculation is
   provided both in blank (dotted line) and lensing fields (solid line) (here
   AC114) and for the three LFs (from right to left, (a) in red, (b) in blue
   and (c) in black)\label{EMIR_mags}} 
\end{figure}

\begin{figure}
 \centering
 \includegraphics[width=9.5cm]{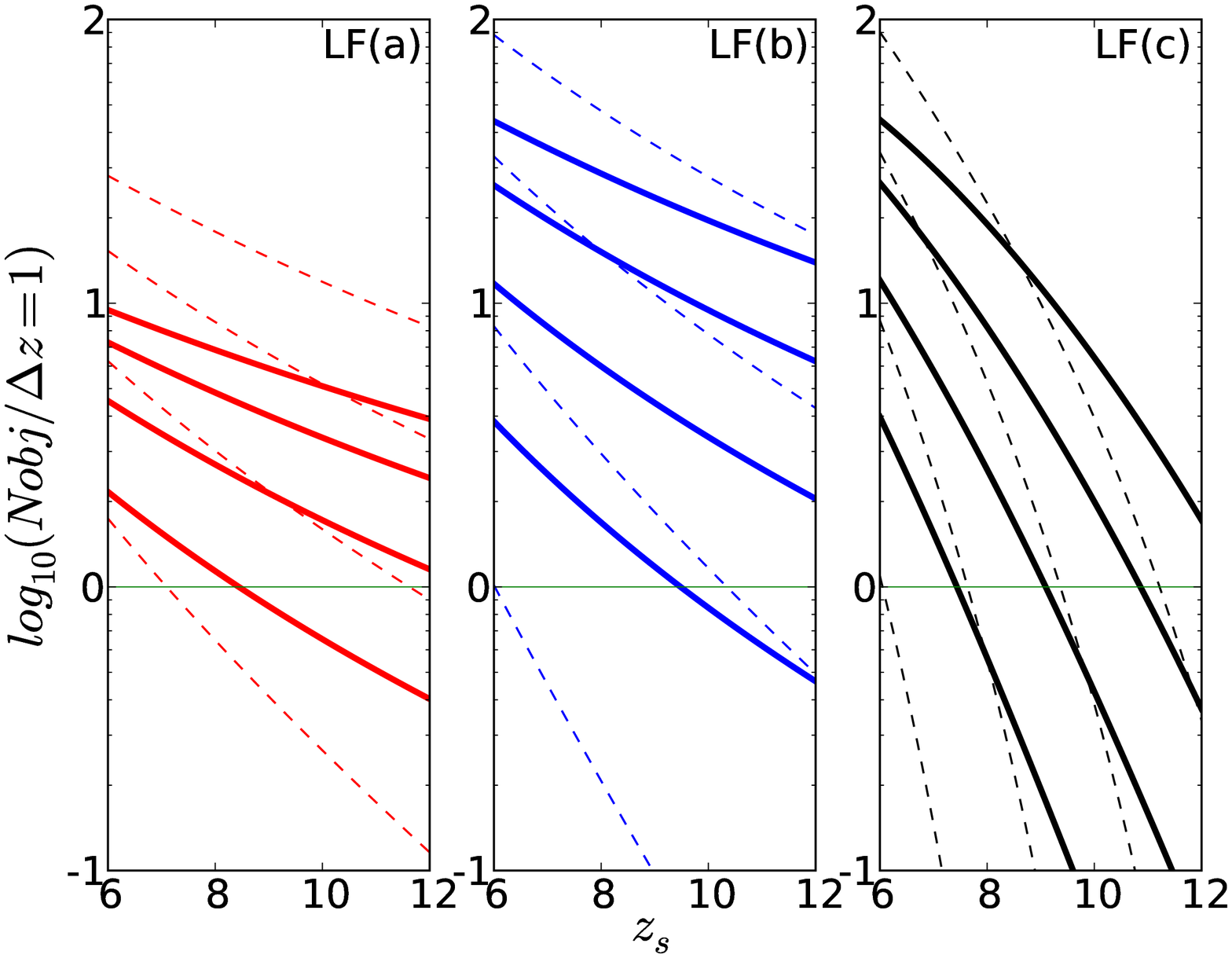}
 \caption{The same as Fig.~\ref{EMIR_mags} but for a $2.2'\times2.2'$
   FOV. Some differences appear in comparison with the
   Fig.~\ref{EMIR_mags}. For example, the total numbers of high-z galaxies
   expected behind lensing clusters (solid lines) and the field (dashed lines)
   are much larger at low limiting magnitude ($m_{AB}=26.0$) but this
   phenomenom is reversed for deeper surveys ($m_{AB}=28.0-29.0$) (see text
   for details). \label{JWST_mags}} 
\end{figure}

\begin{figure}
 \centering
 \psfrag{MAB}[c][c]{$M_{AB} (1500\AA)$}
 \includegraphics[width=9.5cm]{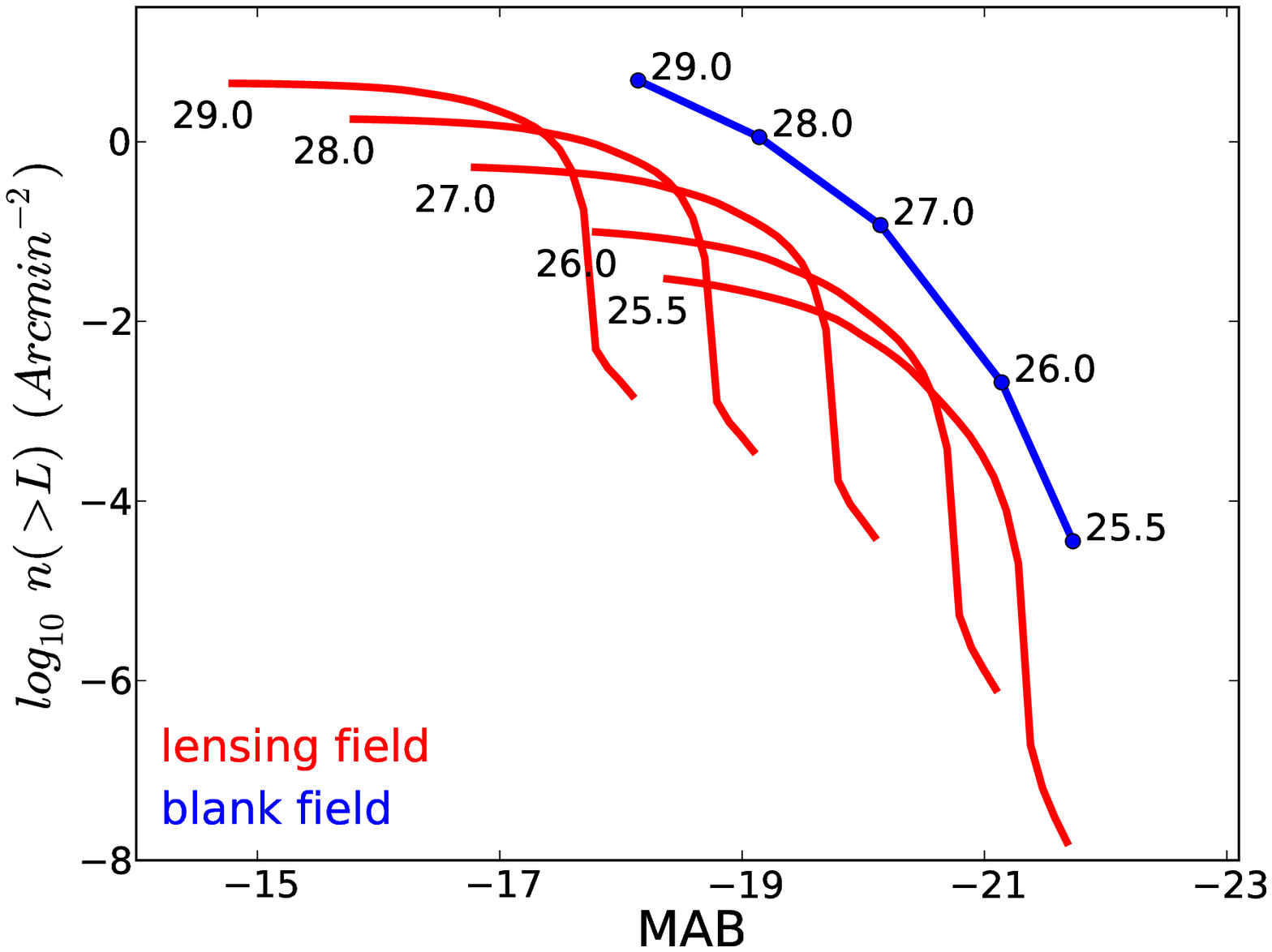}
 \caption{Cumulative surface density of sources as a function of their
   intrinsic UV luminosity, in blank fields (blue solid line) and in the
   lensing fields with FOV $\sim5$ arcmin$^2$ (JWST-like, red solid line), for
   different photometric depths ranging from shallow ($AB\sim25.5$) to deep
   ($AB\sim29.0$) surveys and a strongly evolving LF(c). The source redshift
   is arbitrarily fixed at z=8, with $\Delta z=1$\label{intrinsic_true}} 
\end{figure}

\begin{figure}
 \centering
 \psfrag{SFR}[c][c]{SFR $(\msun/yr)$}
 \psfrag{MAB}[c][c]{$M_{AB} (1500\AA)$}
 \includegraphics[width=9.5cm]{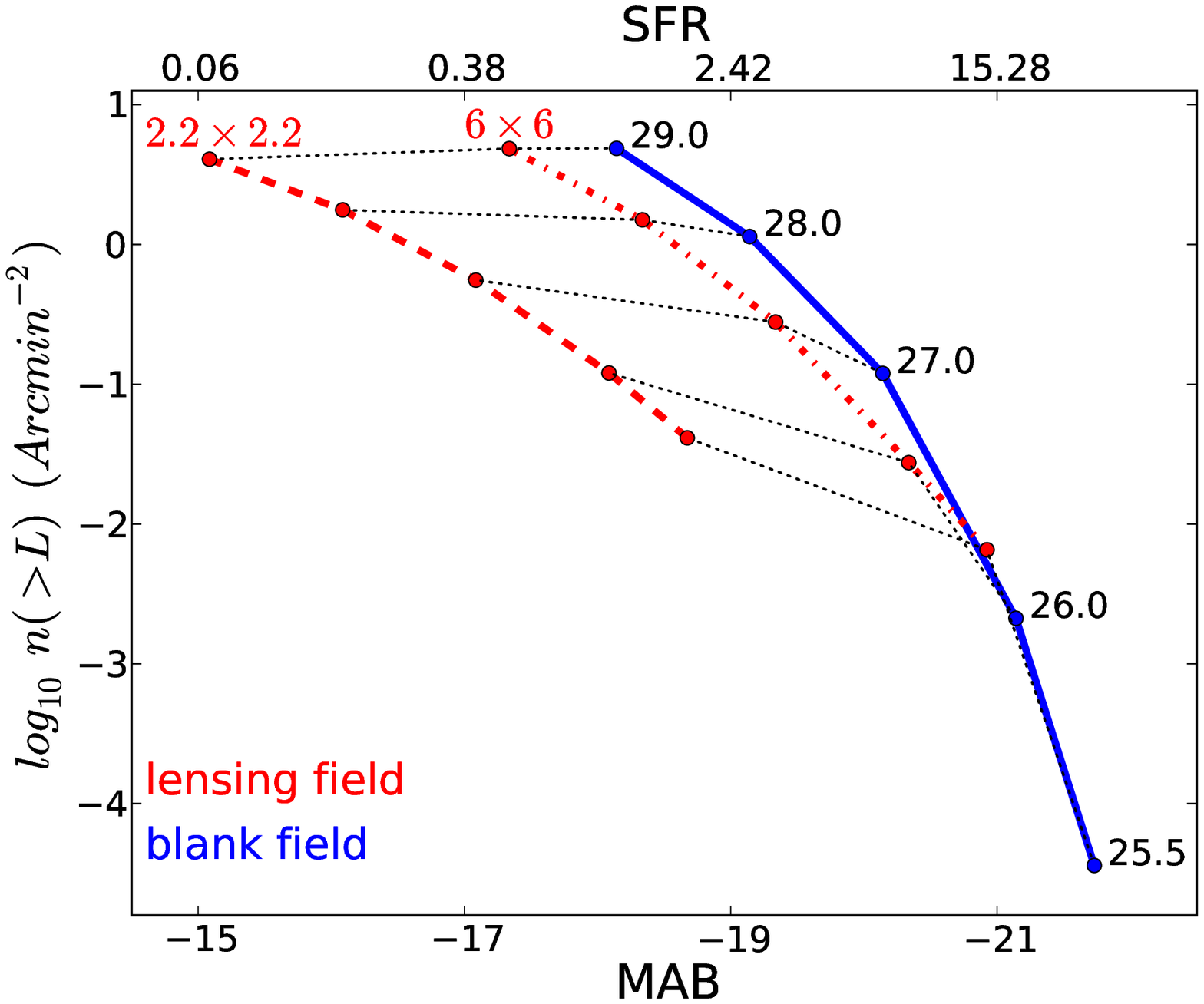}
 \caption{Cumulative surface density of sources as a function of their
   intrinsic UV luminosity and SFR, in blank fields (blue solid line) and in
   the lensing fields with FOV $6'\times6'$ (red dotted line) and
   $2.2'\times2.2'$ (JWST-like, red dashed line), for different photometric
   depths ranging from shallow ($AB\sim25.5$) to deep ($AB\sim29.0$) surveys
   and a strongly evolving LF(c). The mean magnification over the whole the
   field is used to derive the lensing points, the true distribution is
   displayed in Fig.~\ref{intrinsic_true}. The source redshift is arbitrarily
   fixed at z=8, with $\Delta z=1$. The conversion from absolute magnitude to
   SFR is provided in Sect.~\ref{simulation} using the calibrations from
   \citet{Kennicutt}. \label{intrinsic}} 
\end{figure}

\subsection{\label{NB} Lyman Break versus NB searches}

\begin{figure}
 \centering
 \includegraphics[width=9.5cm]{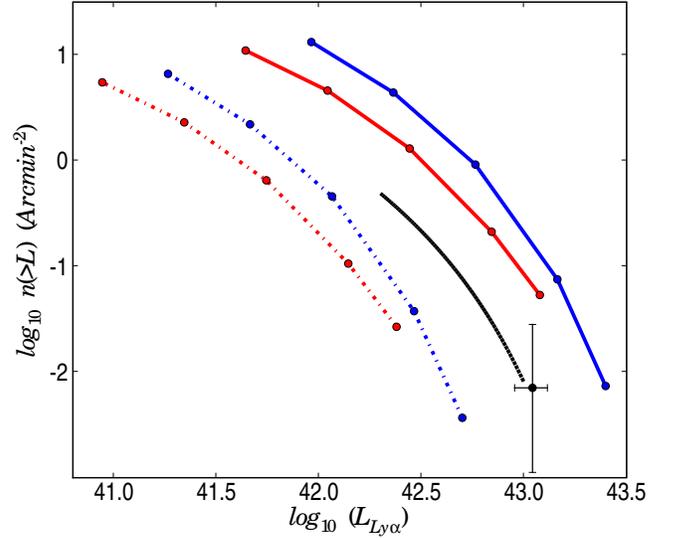}
 \caption{Cumulative surface density of observed sources as a function of
   their Lyman alpha luminosity ($6'\times6'$ FOV, $\Delta_{z}=1$, redshift of
   sources fixed at 6.6 for the LF(c)). The density is calculated in blank
   (blue solid line) and in lensing fields (red solid line) for different
   limiting magnitudes (from right to left: 25.5, 26.0, 27.0, 28.0 and
   29.0). Dashed lines display number counts corrected by transmission value
   $\epsilon \sim10\%$ (dashed red line in lensing fields and dashed blue line
   in blank field). For comparison, raw number counts extracted from the
   spectroscopic sample of LAEs by ~\citet{Kashikawa} are also given (black
   solid line). The number density derived from \citet{Iye06} at $z=6.96$ is
   also indicated, together with corresponding error bars (see text for
   details). As in Fig.~\ref{intrinsic}, the
   magnification used to derive the lensing points is averaged over the
   entire field.
\label{lya_emir}
}

\end{figure}

In this section, we discuss on the relative efficiency of blank and lensing
fields on the detection of LAEs based on either NB surveys or the
spectroscopic follow up of LBGs at z$>$6.  
Although the observational effort required to select $z\ge7$ candidates using
the dropout technique seems relatively cheap as compared to the NB approach,
the two approches are complementary, as emphasized by the fact that many objects found by Lyman
$\alpha$ emission remain weak or undetected in the continuum 
(e.g. \citealt{Rhoads01}, \citealt{Kodaira}, \citealt{Cuby03}, \citealt{Taniguchi}). 
A quantitative comparison between the properties of LAEs and LBGs at $z\ge7$
within the same volume should provide important information on the Lyman
$\alpha$ transmission, SFR and other properties of these high-$z$ galaxies.

Since the pioneering Large Area Lyman Alpha Survey (LALA, \cite{Rhoads01},
\cite{Rhoads03}), different NB surveys in blank fields 
have provided interesting galaxy samples in the z$\sim 5-7$ interval, e.g.
the large sample of \lya \ emitters at $z\sim5.7$ by \citet{Hu04},
the z=6.17 and 6.53 galaxies found respectively by \citet{Cuby03} and
\citet{Rhoads04}, the two $z\sim 6.6$ galaxies detected by \citet{Kodaira},
and the galaxy at a redshift z$=$6.96 found by \citet{Iye06}.
In the latter case, which should be representative of z$\sim7$ samples, 
the authors used a combination of NB imaging at 8150\AA \ (SuprimeCam) 
and broad-band photometry in the optical bands to select candidates for a
subsequent spectroscopic follow up with DEIMOS/Keck. Their confirmation rate is
relatively high (18 sources out of 26 candidates), leading to
0.03 sources/arcmin$^2$ and redshift bin $\delta z=0.1$. Similar results are 
reported by \citet{Kashikawa}. All these sources
have important Lyman $\alpha$ fluxes (a few $\times 10^{-17}$ erg
cm$^{-2}$ s$^{-1}$), and display broad Lyman $\alpha$ lines 
($\sigma_v \sim 200$ km/s). 
A strong evolution is found in the number density of LAEs at $z\ge7$ with
respect to the z$\sim 5-7$ interval (\citealt{Iye06}, \citealt{Willis_ZEN},
\citealt{Cuby07}). 

The number of LAEs expected within a sample of LBGs at $z\gtapprox6$ can be
estimated using the distribution of Lyman-$\alpha$ equivalent widths derived
for the spectroscopic sample of LBGs at $z\sim3$ by \citet{Shapley}, assuming
no evolution in the population of LAEs with respect to LBGs. This simplistic
scaling should be enough for the simulation needs. We introduce a factor
$\epsilon$, defined below, which can be linked to the Lyman $\alpha$
transmission as follows: 
\begin{equation}
\epsilon = \frac{L_{1500}}{L_{\lya}(1+z)} = \frac{1}{W_{\lya}}
\end{equation}
where $L_{1500}$ is the UV monochromatic luminosity at 1500 A, $L_{\lya}$ is
the Lyman-$\alpha$ luminosity and $W_{\lya}$ is the Lyman-$\alpha$ equivalent
width. 
With this simple assumption, the average value for the Lyman-$\alpha$
equivalent width is $\sim10$ $\AA$, corresponding to $\epsilon\sim10\%$. This
value can be used to derive a rough estimate of expected number density of
LAEs, from a population of LBGs 
In addition, the number density is also corrected to take into account of the
fraction of the LBGs sample displaying $\lya$ in emission. 

Fig.~\ref{lya_emir} displays the cumulative number counts of sources at
$z\sim6.6$ integrated from the LF(c) as a function of the Lyman-$\alpha$
luminosity, scaled according to the UV luminosities (cf Sect.~\ref{sources})
in the typical $6'\times6'$ FOV, together with a comparison of observations in
a similar redshift domain ($z_s\sim6.6$ for Kashikawa spectroscopic sample of
LAEs and \citealt{Iye06} at z=6.96).  

The number density of LBGs at z=7 (with $\delta z=0.1$, close to the
band-width of NB surveys) ranges between 0.001 (LF(c)) and 0.02 (LF(a))
sources/$arcmin^2$ for a survey limited to $H(AB)\ltapprox25.5$, depending on
the LF. Lensing clusters improve these numbers by a factor ranging between 6
(for LF(c)) and 2 (for LF(a)). In case of a deep survey limited to
$H(AB)\ltapprox29.0$, the number densities reach 1 (LF(c)) to 2 (LF(a))
sources/$arcmin^2$. In this case, there is a negative magnification bias of
the order of 20\%. These numbers, obtained with a simplistic model, are
between a factor of $\sim10$ (for bright sources) and a few (for faint
sources) smaller than the number densities obtained by ~\citet{Kashikawa} for
their spectroscopic sample. With increasing $z_s$ (see Fig.~\ref{EMIR_mags})
for instance at z=9 with the strongly evolving LF(c), no sources can be
detected for a shallow survey limited to $H(AB)\ltapprox25.5$ and for a deeper
limited survey ($H(AB)\ltapprox29.0$), a minimum of 3 $arcmin^2$ surveyed area
is needed to obtain 1 source in a blank field. In a lensing field with the
(LF(c)), these number densities reach 0.002 for $H(AB)\ltapprox25.5$ and 0.32
sources/$arcmin^2$ for $H(AB)\ltapprox29.0$. The relatively low-efficiency of
lensing clusters with NB techniques in the $z\ge9$ domain has been recently
confirmed by the results of \citet{Willis_ZEN3}.  

The preselection of $z\ge6-7$ candidates in lensing fields has two main
advantages with respect to blank fields. In the shallow ($AB\le25.5$) regime,
there is an increase by a factor $\sim8-10$ on the number of sources detected
and a moderate gain in depth for a given exposure time (i.e. $\sim0.5$
magnitudes at $AB\sim25.5$). In the deep-survey regime ($AB\le28-29$), there
is a gain in intrinsic depth, for a number of candidates which remains
essentially constant (i.e. $\sim0.8$ mags gain at $AB\le28$). The relative
efficiency of lensing with respect to blank field counts in
Fig.~\ref{intrinsic} depends on the FOV. The two predictions get close to each
other with increasing values of the FOV in lensing surveys, and the trend goes
in the opposite direction for smaller FOV. This trend is the same for both
LBGs and LAEs. To explore the bright end of the LF, blank field surveys are
needed with a large FOV, whereas lensing clusters are particularly useful to
explore the faint end of the LF. This trend is further discussed in the next
Section.  

\subsection{\label{ideal} Towards the ideal survey: constraining the Luminosity Function of high-$z$ sources}

All present photometric surveys aimed at constraining the UV LF at z$>$7,
either space or ground-based, are still dramatically small in 
terms of effective surface. Wide and deep optical$+$
near-IR surveys in lensing and blank fields are needed to set strong
constraints on the LF and on the star-formation density at
z$>$7.  An important issue is the combination between photometric depth and
surveyed area which is needed to identify a representative number of
photometric candidates, or to reach a significant non-detection limit in order
to constrain the LF of z$>$7 sources. 

There are three different aspects to consider when designing an ``ideal''
survey aiming at constraining the LF: the depth and the area of the survey,
and the corresponding field to field variance. In order to address these
issues, we have computed the expected field to field variance corresponding to
lensing and blank field surveys, for different survey configurations (area and
depth). A summary of these results is given in Table~\ref{tab_vr_nfields} for
different number of lensing clusters, and for two representative depths in the
H-band (i.e. a ``shallow'' survey with $AB\le25.5$, and a ``deep'' survey with
$AB\le28.0$) assuming a strongly evolving LF(c) in all cases. This table
complements the results given in Tables~\ref{tab_vr} and~\ref{tab_vr_allmag}
for blank and lensing fields as a function of depth. In all cases, we use
AC114 as a reference for lensing clusters. 

Regarding field-to-field variance in number counts, results are expected to be
similar in blank and lensing fields for a relatively wide FOV ($\sim40-50$
$arcmin^2$; see Sect.~\ref{variance} and Table~\ref{tab_vr_allmag}). As shown
in Table~\ref{tab_vr_nfields}, a deep lensing survey using $\sim10$ clusters
should be able to reach a variance $\ltapprox20$\% on sources at $6\le z
\le8$, irrespective of the actual LF. This value is better than present-day
photometric surveys in blank fields, typically reaching 30-35\% for
$AB\ltapprox29.0$ (e.g.~\citealt{Bouwens08}), which in turn is rather close to
what could be achieved in a single lensing cluster for $AB\ltapprox28.0$. 

A different survey strategy consists of increasing the number of lensing
clusters with a shallow limiting magnitude. In this case, a few tens of
lensing clusters (typically between 10 and 50, depending on the LF) are needed
to reach a variance of $\sim30$\% at $z\ltapprox8$. Note that the difference
in exposure time between the shallow and deep surveys reported in
Table~\ref{tab_vr_nfields} is a factor $\sim100$, and that $\sim10$ pointings
are needed on blank fields in order to reach the same number of $z\sim6-8$
sources as in a single ``shallow'' lensing field. 

In the case of a strongly evolving LF(c), photometric surveys should reach a
minimum depth of $AB\sim28$ to achieve fair statistics on $z\sim9-10$ sources
using a lensing cluster (Table~\ref{tab_vr_allmag}). In this case we expect
between 20 (z=9, $\sim40$\% variance) and 8 (z=10, $\sim40$\% variance)
sources per lensing cluster in a $\sim40$ $arcmin^2$ FOV. The efficiency is a
factor of 10 smaller at $z\sim10$ in blank fields. Fair statistics at
$z\sim12$ should require a minimum depth close to $AB\sim29$ both in lensing
and blank fields. 

Constraining the LF of star forming galaxies at $z\ge7$ should require the
combination of blank and lensing field observations. This is illustrated for
instance in Fig.~\ref{intrinsic_true} and ~\ref{intrinsic} for an example at
$z_s=8$. A survey towards a lensing cluster has several advantages. It
increases the total number of sources available for spectroscopic follow up,
and it helps extending the sample towards the faint edge of the LF and towards
the highest possible limits in redshift. On the other hand, blank fields are
needed to achieve fair statistics on the bright edge of the LF. Thus an
``ideal'' survey should combine both blank and lensing fields. Given the
numbers presented in previous sections, a blank field used for these purposes
should be a factor ranging between $\sim10$ and $100$ times larger than a
lensing field (depending on the redshift domain, photometric depth, and actual
LF) in order to efficiently complete the survey towards $L>L^*$. This should
be possible with the new upcoming surveys, such as the WIRCAM ultra deep
survey (WUDS) at CFHT ($\sim400$ arcmin$^2$, with $YJHK\ltapprox25.5$),
UKIDSS-UDS ($\sim2700$ arcmin$^2$, with $YJHK\ltapprox25$) or Ultra-Vista
($\sim2600$ arcmin$^2$, with $YJH\ltapprox26$, $K\ltapprox25.6$). The optimum
number of lensing fields ranges between $\sim10-20$ (for $z\sim6-8$ studies
with ``shallow'' photometry) to a few (for ``deep'' surveys targeting
$z\sim8-12$ sources). 

\begin{table*}
\centering 
\caption{Field to field variance, 
including $1-\sigma$ errors on the magnification factor, 
  expected in a lensing survey, as a function of the number of
  clusters, for the three LFs (a), (b) and (c) (from top to bottom
  respectively). Simulations were performed using the same approaches in
  Sect.~\ref{variance} (i.e. $6'\times6'$ FOV, ...) for two different depths:
  $AB\le 25.5$ (shallow survey) and $AB\le 28.0$ (deep
  survey) \label{tab_vr_nfields}} 

\begin{tabular}{c|c c c c c c} 
$v_r$ LF(a)/$N_{clusters}$ & 1 & 6 & 10 & 20 & 50 & 100 \\
\hline
$z=6$ \\
\\
Shallow  & $50^{+3}_{-3}$\% & $28^{+1}_{-1}$\% & $24^{+1}_{-1}$\% & $19^{+0}_{-0}$\% & $14^{+0}_{-0}$\% & $10^{+0}_{-0}$\%  \\
\\
Deep & $25^{+1}_{-1}$\% & $17^{+0}_{-0}$\% & $15^{+0}_{-0}$\% & $12^{+0}_{-0}$\% & $9^{+0}_{-0}$\% & $7^{+0}_{-0}$\%  \\
\\
\hline
$z=7$ \\
\\
Shallow & $62^{+2}_{-2}$\% & $33^{+1}_{-1}$\% & $28^{+1}_{-1}$\% & $22^{+0}_{-0}$\% & $16^{+0}_{-0}$\% & $12^{+0}_{-0}$\% \\
\\
Deep & $29^{+1}_{-1}$\% & $19^{+0}_{-0}$\% & $17^{+0}_{-0}$\% & $13^{+0}_{-0}$\% & $10^{+0}_{-0}$\% & $7^{+0}_{-0}$\% \\
\\
\hline
$z=8$ \\
\\
Shallow & $75^{+3}_{-3}$\% & $39^{+1}_{-2}$\% & $32^{+1}_{-1}$\% & $25^{+0}_{-1}$\% & $18^{+0}_{-0}$\% & $13^{+0}_{-0}$\% \\
\\
Deep & $32^{+1}_{-1}$\% & $21^{+0}_{-0}$\% & $18^{+0}_{-0}$\% & $15^{+0}_{-0}$\% & $11^{+0}_{-0}$\% & $8^{+0}_{-0}$\% \\

\end{tabular}
\end{table*}

\begin{table*}
\centering 
\begin{tabular}{c|c c c c c c} 
$v_r$ LF(b)/$N_{clusters}$ & 1 & 6 & 10 & 20 & 50 & 100 \\
\hline
$z=6$ \\
\\
Shallow & $57^{+4}_{-4}$\% & $40^{+2}_{-2}$\% & $33^{+2}_{-2}$\% & $26^{+1}_{-1}$\% & $18^{+0}_{-0}$\% & $14^{+0}_{-0}$\%  \\
\\
Deep & $23^{+1}_{-1}$\% & $14^{+0}_{-0}$\% & $13^{+0}_{-0}$\% & $10^{+0}_{-0}$\% & $8^{+0}_{-0}$\% & $6^{+0}_{-0}$\%  \\
\\
\hline
$z=7$ \\
\\
Shallow & $77^{+6}_{-6}$\% & $40^{+2}_{-2}$\% & $33^{+1}_{-1}$\% & $25^{+0}_{-1}$\% & $18^{+0}_{-0}$\% & $13^{+0}_{-0}$\% \\
\\
Deep & $24^{+1}_{-1}$\% & $17^{+0}_{-0}$\% & $15^{+0}_{-0}$\% & $12^{+0}_{-0}$\% & $9^{+0}_{-0}$\% & $7^{+0}_{-0}$\% \\
\\
\hline
$z=8$ \\
\\
Shallow & $100^{+8}_{-9}$\% & $48^{+2}_{-3}$\% & $40^{+2}_{-2}$\% & $30^{+1}_{-1}$\% & $21^{+0}_{-0}$\% & $15^{+0}_{-0}$\% \\
\\
Deep & $28^{+1}_{-1}$\% & $19^{+0}_{-0}$\% & $16^{+0}_{-0}$\% & $13^{+0}_{-0}$\% & $10^{+0}_{-0}$\% & $7^{+0}_{-0}$\% \\

\end{tabular}

\end{table*}

\begin{table*}
\centering 
\begin{tabular}{c|c c c c c c} 
$v_r$ LF(c)/$N_{clusters}$ & 1 & 6 & 10 & 20 & 50 & 100 \\
\hline
$z=6$ \\
\\
Shallow & $57^{+4}_{-4}$\% & $40^{+2}_{-2}$\% & $33^{+1}_{-1}$\% & $26^{+1}_{-1}$\% & $18^{+0}_{-0}$\% & $14^{+0}_{-0}$\%  \\
\\
Deep & $23^{+0}_{-0}$\% & $14^{+0}_{-0}$\% & $13^{+0}_{-0}$\% & $10^{+0}_{-0}$\% & $8^{+0}_{-0}$\% & $6^{+0}_{-0}$\%  \\
\\
\hline
$z=7$ \\
\\
Shallow & $108^{+12}_{-10}$\% & $51^{+3}_{-4}$\% & $42^{+2}_{-2}$\% & $32^{+1}_{-1}$\% & $22{+1}_{-1}$\% & $16^{+0}_{-0}$\% \\
\\
Deep & $26^{+1}_{-1}$\% & $18^{+0}_{-0}$\% & $15^{+0}_{-0}$\% & $12^{+0}_{-0}$\% & $9^{+0}_{-0}$\% & $7^{+0}_{-0}$\% \\
\\
\hline
$z=8$ \\
\\
Shallow & $211^{+38}_{-29}$\% & $92^{+8}_{-8}$\% & $73^{+6}_{-6}$\% & $53^{+4}_{-4}$\% & $35^{+1}_{-1}$\% & $25^{+1}_{-1}$\% \\
\\
Deep & $31^{+1}_{-1}$\% & $21^{+0}_{-0}$\% & $18^{+0}_{-0}$\% & $15^{+0}_{-0}$\% & $11^{+0}_{-0}$\% & $8^{+0}_{-0}$\% \\

\end{tabular}

\end{table*}


\section{\label{conclusions} Summary and conclusions}

We have evaluated the relative efficiency of lensing clusters with respect to
blank fields in the identification and study of $z\ge6$ galaxies. The main
conclusions of this study are given below.  

For magnitude-limited samples of LBGs at $z\ge6$, the magnification bias
increases with the redshift of sources and decreases with both the depth of
the survey and the size of the surveyed area. Given 
the typical near-IR FOV 
in lensing fields, the maximum efficiency is reached for clusters at
$z\sim0.1-0.3$, with maximum cluster-to-cluster differences ranging between 30
and 50\% in number counts, depending on the redshift of sources and the LF. 

The relative efficiency of lensing with respect to blank fields strongly
depends on the shape of the LF, for a given photometric depth and FOV. The
comparison between lensing and blank field number counts is likely to yield
strong constraints on the LF. 

The presence of a strong-lensing cluster along the line of sight has a
dramatic effect on the observed number of sources, with a positive
magnification effect in typical ground-based ``shallow'' surveys
($AB\le25.5$). The postive magnification bias increases with the redshift of
sources, and also from optimistic to pessimistic values of the LF. In case of
a strongly evolving LF at $z\ge7$, as proposed by \cite{Bouwens08}, blank
fields are particularly inefficient as compared to lensing fields. For
instance, the size of the surveyed area in ground-based observations would
need to increase by a factor of $\sim10$ in blank fields with respect to a
typical $\sim30-40$ $arcmin^{2}$ survey in a lensing field, in order to reach
the same number of detections at $z\sim6-8$, and this merit factor increases
with redshift. 
All these results have been obtained assuming that number counts derived
in clusters are not dominated by sources below the limiting surface brightness
of observations, which in turn depends on the reliability of the usual
scalings applied to the size of high-z sources. 

Ground-based ``shallow'' surveys are dominated by field-to-field variance
reaching $\sim$ 30 to 50\% in number counts between z$\sim$6 and 8 in a unique
$\sim30-40$ $arcmin^{2}$ lensing field survey (or in a 400 $arcmin^{2}$ blank
field), assuming a strongly evolving LF. 

The number of z$>$8 sources expected at the typical depth of JWST
($AB\sim28-29$) is much higher in lensing than in blank fields if the UV LF is
rapidly evolving with redshift (i.e. a factor of $\sim10$ at $z\sim10$ with
$AB\ltapprox28$). 

Blank field surveys with a large FOV are needed to probe the bright edge of
the LF at $z\ge6-7$, whereas lensing clusters are particularly useful to
explore the mid to faint end of the LF. 


\begin{acknowledgements}

We are grateful to D. Schaerer, A. Hempel, J.F. Le Borgne and E. Egami for
useful comments. We acknowledge financial support from the European
Commissions ALFA-II programme through its funding of the Latin-America
European Network for Astrophysics and Cosmology (LENAC). This work was also
supported by the French {\it Centre National de la Recherche Scientifique},
the French {\it Programme National de Cosmologie} (PNC) and {\it Programme
  National de Galaxies} (PNG). JR acknowledges support from a EU Marie-Curie
fellowship. 

\end{acknowledgements}


\bibliography{roserbib}


\end{document}